\documentclass[11pt]{article}
\usepackage{acl}

\newif\ifdraft\draftfalse            %
\newif\ifanon\anonfalse              %

\usepackage[T1]{fontenc}
\usepackage[utf8]{inputenc}
\usepackage{times}
\usepackage{latexsym}
\usepackage{microtype}
\usepackage{inconsolata}
\usepackage{graphicx}
\usepackage{textcomp}
\usepackage{xspace}
\usepackage{cleveref}
\usepackage{array}
\usepackage{booktabs} % For \toprule, etc.
\usepackage{listings}
\usepackage{siunitx} % For \num
\usepackage[normalem]{ulem} % for an underline that line-breaks
\usepackage{tikz}
\usetikzlibrary{arrows.meta,positioning}
\usepackage{circledsteps}
\usepackage{subcaption} %for last figure
\usepackage{makecell}
\usepackage{colortbl} %for significance tables
\usepackage{multirow} % For \multirow in tables
\usepackage{enumitem} % For description environment options
\usepackage{mdframed}
\usepackage{amsfonts} % for mathbb
\usepackage{amssymb} % for graph symbols
\usepackage{xparse}
\usepackage{cuted}

\newenvironment{epigraph}%
  {\setlength{\topsep}{0pt}\setlength{\partopsep}{0pt}\begin{quote}}%
  {\end{quote}}
\makeatletter
\renewcommand{\verbatim@font}{\rmfamily}
\makeatother

\Crefname{section}{\S}{\S\S}
\crefformat{section}{\S#2#1#3}
\Crefname{figure}{Figure}{Figures}
\crefformat{figure}{Figure #2#1#3}
\Crefname{Figure}{Figure}{Figures}
\Crefname{Table}{Table}{Tables}
\crefformat{table}{Table #2#1#3}

\definecolor{lilac}{RGB}{200, 162, 200}
\definecolor{wellesleyblue}{RGB}{0, 39, 118}
\definecolor{brownbrown}{RGB}{78,54,41}
\definecolor{northeasternred}{RGB}{200, 16, 46}

\newcommand{\qwenFull}{\textsc{Qwen3.6-35B-A3B}\xspace}
\newcommand{\gemmaFull}{\textsc{Gemma-4-26B-A4B}\xspace}
\newcommand{\sonnetFull}{\textsc{Claude-Sonnet-4.6}\xspace}
\newcommand{\gptFull}{\textsc{GPT-5.5}\xspace}
\newcommand{\glmFull}{\textsc{GLM-5.2}\xspace}
\newcommand{\qwen}{Qwen\xspace}
\newcommand{\gemma}{Gemma\xspace}
\newcommand{\sonnet}{Sonnet\xspace}
\newcommand{\gpt}{GPT-5.5\xspace}
\newcommand{\glm}{GLM\xspace}

\newcounter{trajcounter}

\newcommand{\transcriptturn}[1]{%
  \par\smallskip\noindent\textbf{#1:}\enspace\ignorespaces%
}

\lstdefinestyle{codeblock}{
    basicstyle=\ttfamily\footnotesize,
    frame=none,
    backgroundcolor=\color{white},
    commentstyle=\color{red!60!black},
    keywordstyle=\color{green!50!black},
    stringstyle=\color{red!60!black},
    basicstyle=\ttfamily\footnotesize,
    breakatwhitespace=false,         
    breaklines=true,                 
    captionpos=b,                    
    keepspaces=true,                 
    showspaces=false,                
    showstringspaces=false,
    showtabs=false,                  
    tabsize=2,
    escapechar={~},
}

\NewDocumentEnvironment{transcript}{o}{%
  \refstepcounter{trajcounter}%
  \IfValueT{#1}{\label{#1}}%
  \par\smallskip
  \begingroup
  \footnotesize
  \setlength{\parskip}{0.25\baselineskip}%
  \def\system{\transcriptturn{System}}%
  \def\assistant{\transcriptturn{Assistant}}%
  \def\user{\transcriptturn{User}}%
  \lstset{%
    basicstyle=\ttfamily\scriptsize,
    breaklines=true,
    columns=fullflexible,
    keepspaces=true,
    showstringspaces=false,
    frame=none,
    aboveskip=0.25\baselineskip,
    belowskip=0.25\baselineskip
  }%
  \begin{mdframed}[
    backgroundcolor=gray!10,
    linecolor=gray!10,
    linewidth=0pt,
    innertopmargin=0.5\baselineskip,
    innerbottommargin=0.5\baselineskip,
    innerleftmargin=0.75em,
    innerrightmargin=0.75em,
    skipabove=0.75\baselineskip,
    skipbelow=0.75\baselineskip
  ]%
}{%
  \end{mdframed}%
  \endgroup
}

\title{The Best Programming Language for Tokenmaxxing\\
\large An Investigation of Coding Agent Behavior Across Programming Languages}

\author{
  Zixuan Wu \\
  Northeastern University
  \And
  Carolyn Jane Anderson \\
  Wellesley College
  \And
  Arjun Guha \\
  Northeastern University
  }

\begin{document}

\maketitle
\begin{strip}
\begin{epigraph}
Let's say you have a software engineer or AI researcher and you pay them
\$500,000 a year. We do that all the time. That \$500,000 engineer, at the end
of the year, I'm going to ask them, how much did you spend in tokens? If that
person said, \$5,000, I will go ape... something else. If that \$500,000
engineer did not consume at least \$250,000 worth of tokens, I'm going to be
deeply alarmed. --- Jensen Huang \citep{allin-huang-2026}
\end{epigraph}
\end{strip}

\begin{abstract}
Although coding agents are now very effective in a variety of programming languages, this paper shows that the cost (in tokens) can vary significantly by programming language. We evaluate five recent models on programming problems in Python, Java, Rust, and OCaml. We carefully control for problem difficulty, and show that there can be stark variation in token consumption that is consistent across models.
To understand why, we analyze both the structure and content of agent trajectories. First, we re-execute every intermediate solution and abstract each trajectory as a sequence of test-outcome vectors, then label the work between successive solutions. This reveals that agents repeatedly produce noncompiling solutions in unfamiliar languages and revise solutions that already pass. Second, we analyze trajectory text, finding that agents plan solutions in code comments, distrust the provided tests in favor of inputs they invent, and sidestep unfamiliar target languages by prototyping in Python.

Our results show that by-language token efficiency is a metric that should be considered when benchmarking and developing multilingual agents, and, for the tokenmaxxer, provide a guide to the most expensive language.

\end{abstract}

%% ============================================================

% \begin{quote}
% Let's say you have a software engineer or AI researcher and you pay them
% \$500,000 a year. We do that all the time. That \$500,000 engineer, at the end
% of the year, I'm going to ask them, how much did you spend in tokens? If that
% person said, \$5,000, I will go ape... something else. If that \$500,000
% engineer did not consume at least \$250,000 worth of tokens, I'm going to be
% deeply alarmed. --- Jensen Huang \citep{allin-huang-2026}
% \end{quote}

\section{Introduction}
\label{sec:intro}

Over the past 1\textonehalf{} years, coding agents have gone from research
idea~\citep{yang2024sweagent} to tools that are widely used by programmers.
Although many programmers use agents by choice, others feel compelled to do so
to meet corporate targets for AI adoption. There are even reports of
organizations running internal leaderboards for token consumption. Following
Goodhart's Law~\citep{Strathern_1997}, some have resorted to
\emph{tokenmaxxing}, or giving agents tasks for the purpose of
increasing LLM spending, without regard to whether it is worth the cost.  We,
the authors of this paper, think this is definitely
pointless and possibly unethical. But, for those who must tokenmax, we present a
principled way to do so:  work in the programming language that makes the agent
consume the most tokens. Naively, one might think that this is the language in
which programs are the longest, but this isn't the case with agents that reason
and self-correct. What matters is not the number of tokens in the final
solution, but the amount of reasoning and revision needed to get to the
solution.

More seriously, the goal of this paper is to study how the behavior of
software-engineering agents varies by programming language. We evaluate four
programming languages (Java, OCaml, Python, and Rust) and five models on a
100-problem subset of LiveCodeBench, gathering 2{,}000 agent trajectories for
analysis.

Our basic finding is that the choice of programming language has a significant
effect on output tokens produced by the agent, even after carefully controlling
for problem difficulty. The effect is most pronounced for OCaml across all
models, but is significant for Rust and Java too. Models are most efficient when
they work in Python.

However, we go beyond counting tokens and analyze agent trajectories in several
ways. First, we evaluate every intermediate solution that the agents produce,
which allows us to study an agent's progression on the unit tests for a problem.
This allows us to observe and quantify several behaviors: agents repeatedly
struggling with the same compile error; agents that get stuck revising a
solution even when all tests pass; and so forth. Second, we label the actions in
every trajectory with an LLM using a 12-category taxonomy that we develop. This
allows us to ask what an agent is doing as it works through a problem. These
labels allow us to determine the turns spent on bug fixes, optimizations,
exploration, refactors, and more. Finally, we analyze the text the agents
produce as they work, which reveals specific behaviors: agents sometimes plan in
comments before writing code; they sometimes second-guess and re-derive
already-passing solutions; they can distrust the provided tests in favor of
inputs they invent; and they can sidestep an unfamiliar language by prototyping
in Python first.

\paragraph{Contributions} We make the following contributions.
  (1)~We study how token cost in agentic coding varies
  across programming languages, controlling for problem difficulty. (2)~We
  find that, at comparable success rates, lower-resource languages cost
  significantly more than Python. (3)~We introduce a novel analysis
  approach that involves testing every intermediate solution that an agent writes in an external test harness, which allows us to analyze agents' test progression. (4)~We
  classify the steps between intermediate solutions as semantic
  actions, revealing specific agent behaviors and where they fall on the path to
  a solution.

These findings show how the choice of programming language shapes
an agent's token cost and which behaviors drive the difference between
languages, offering insight toward building more efficient agents across
programming languages. We will release all code and data upon publication.

\section{Background and Related Work}
\label{sec:background}

\subsection{Multilingual Code Evaluation}

Many benchmarks evaluate how the choice of programming language affects models'
code generation performance. The earliest benchmarks translated single-function
code completion tasks into other
languages~\citep{multipl_e,athiwaratkun2022multi}.  Multi-SWE-bench and SWE-bench
Multilingual \citep{zan2026multi,yang2025swesmith} evaluate agents on their
ability to solve real-world issues in large code repositories. The tasks that we
study lie between code completion and repository-level tasks. We do not
introduce a new benchmark, but use LiveCodeBench in a multi-language
context to study agent trajectories.

\subsection{Agentic Coding and Process Cost}

Beyond correctness, prior work measures several notions of code-generation
cost. One body of work studies the efficiency of the generated program itself---its
runtime and memory across languages~\citep{fan2025sweeffi,qing2025effibenchx}.
Another studies the token cost of the program's surface representation, where
formatting and verbose recurring patterns add tokens without functional
benefit~\citep{pan2026hiddencost,sun2025tokensugar}. In an agentic setting,
however, multi-turn reasoning and tool use can cost far more than the final
program. Existing work on this process cost focuses almost entirely on
Python~\citep{bai2026ai}. We instead make the target language the independent
variable: process cost is not surface cost, and the language in which an agent
spends the most tokens need not be the one whose programs are longest.

Prior work documents particular sources of agent inefficiency, including
continuing to generate after finding a passing solution
\citep{solovyeva2026babbling} and looping when stuck
\citep{fan2025sweeffi}. Related methods stop trajectories early or prune
redundant work~\citep{guo2026eet,gao2026morewithless,xiao2026reducing,yang2025codeagents}.
On less familiar languages, agents may instead prototype and debug in Python
before translating to the target language~\citep{sharma2026metaprogramming}.
Recent analyses open up the trajectory by decomposing token cost by development
phase~\citep{salim2026tokenomics} or agent action
\citep{bouzenia2025understanding,liu2026process}. Our analysis likewise resolves
a trajectory into phases, but grounds those phases in changes in externally
verified test outcomes between successive program snapshots. This lets us
localize behaviors---including stalling, reworking a passing solution, and
sidestepping through Python---to particular stages of solution progress.

%% ============================================================
\section{Methods}
\label{sec:method}

We first present how we set up the multi-language environment, the models we use, how we prompt agents, and how we run our experiments.

\begin{table*}
\centering
\caption{We run agents in a container based on Ubuntu 24.04. The container has 
the language versions shown below, along with optional packages that may help the
agents.  We build and run agents' solutions with the
compile and run commands shown below, and we provide these commands to the agent in the prompt as part of the task.}
\label{tab:environment}
\small
\begin{tabular}{lllll}
\toprule
Language & Version & Extra Packages & Compile command & Run command \\
\midrule
Python & 3.12.3 &
numpy, pandas, scipy, sympy &
--- &
\texttt{python3 solution.py} \\
OCaml & 5.4.1 &
core, base, zarith &
\texttt{dune build ./solution.exe} &
\texttt{dune exec ./solution.exe} \\
Java (OpenJDK) & 21.0.11 &
--- &
\texttt{javac Solution.java} &
\texttt{java Solution} \\
Rust & 1.96 &
--- &
\texttt{cargo build -{}--release} &
\texttt{./target/release/solution} \\
\bottomrule
\end{tabular}
\end{table*}

\subsection{Choice of Agents, Models, and Languages}

We use mini-swe-agent~\citep{yang2024sweagent}, a customizable agent
that supports any model and has a stable interface. It is also an established
baseline agent that is widely
used~\citep{merrill2026terminal,zhou2026featurebench,li2026contextbench,zhang2026sweexplore}.

We work with five models: \qwenFull, \gemmaFull, \glmFull, \gptFull, and \sonnetFull~\citep{qwen36_35b_a3b,gemma4_google_2026,glm5team2026glm5vibecodingagentic,anthropic_sonnet46_system_card_2026,openai_gpt55_system_card_2026}. We run the open-weight \qwen and \gemma models locally and access the open-weight \glm model via the Hugging Face Inference API; all three are amenable to reproducibility. We also evaluate the proprietary \gpt and \sonnet models that power Codex and Claude Code, respectively. We enable reasoning for all models. Additional implementation details and hyperparameters are provided in \cref{appendix:hyperparams}.

We test agents in Python, Java, Rust, and OCaml. Python is the language used
most widely to benchmark models' programming abilities. Although Java is not
popular in ML research, it remains widely used. Rust and OCaml help illuminate agent behavior on languages that are
not as well-represented in pretraining data.\footnote{We hypothesize that the
pretraining corpora of proprietary models have a distribution similar to the few
public pretraining code corpora available for inspection, such as The Stack
v2~\citep{lozhkov:starcoder2}. After deduplication, The Stack v2 has around
200GB of Java and of Python, but only 12GB of Rust and even less OCaml.}

% \subsection{Measuring Agents' Token Usage}

% For any nontrivial problem that takes more than one turn to solve, the agent's
% conversation history is fed back into the model, turn after turn. So output
% tokens become input tokens, and the total number of input tokens over several
% turns can grow enormous. However, on a good inference platform, input processing
% is much cheaper than generating outputs, especially when the platform supports
% prefix caching~\citep{zheng2024sglang}. When an agent works autonomously to
% solve a problem over several turns, prefix caching is very effective, and what
% matters most is the time required to generate outputs. This is why, like many
% others, we report the total number of output tokens generated over a single
% conversation.

% With contemporary reasoning models, the output tokens include both the reasoning
% and normal output. In this paper, we work with both closed and open-weight and
% reasoning models. The closed models that we use, from OpenAI and Anthropic, do
% not expose the reasoning output to users, but users are still charged for the
% reasoning tokens produced. From the perspective of the user, output tokens cost
% the same, whether or not the output is visible to the user.

\subsection{Environment Configuration}

We run agents in a containerized environment that includes the standard
toolchain for each programming language, and common packages
(\cref{tab:environment}). \cref{tab:environment} also shows how we compile and
run solutions to verify agents' results. For OCaml and Rust, the environment
includes a working configuration for Dune and Cargo, so that the agent does not
need to spend turns configuring the compiler. 
% The agent is thus free to edit the
% Dune/Cargo configuration, or write a solution that spans several files.

At runtime, we limit the agent to 40 turns and 10 minutes of execution for each
problem. We also impose a 256MB memory limit on each container, and as described
below, take steps to prevent code execution from causing timeouts.
\Cref{appendix:hyperparams} describes the host machine configuration.

\subsection{The MiniLCB Benchmark}
\label{sec:setup}
% NOTE(arjun): I'm excluding the statement below. But, for later, I think we should confirm that this isn't a bug that potentially affected Agnostics. More details please. :)
% We exclude problems whose statements accept multiple valid outputs, which our harness cannot reliably grade; 3 of 103 initially drawn problems were removed on this basis, and 3 backfill problems were drawn to maintain a pool of exactly 100. \aw{this would probably affected agnostics.}

LiveCodeBench~\citep{jain2025livecodebench} consists of problems from
programming contest websites and is typically used to evaluate models in a
single attempt. Version 6 contains 880 problems, of which 381 have Python
starter code; the remaining 499 are effectively language-neutral and test
solutions over standard I/O. Agnostics~\citep{boruchgruszecki:agnostics} and
MultiLCB~\citep{ivanova:multilcb} use this observation to evaluate models across
languages with per-language compile-and-run configurations. We rely on the same
observation to construct an agent benchmark.

With five models, four programming languages, and these 499 language-agnostic
tasks, we have the opportunity to run 9{,}980 agent sessions. To conserve tokens
and budget, we instead randomly sample 100 tasks (yielding 2{,}000 sessions),
and we refer to this subset as \textbf{MiniLCB}.  Test cases per problem range
from 2 to 45, with a median of 22. 

In standard LiveCodeBench evaluation, test cases run in an external harness to
validate a model's single solution offline. In MiniLCB, \emph{we make all tests
and the test harness visible to the agent}. Specifically, we present each
problem as a small repository:  a \texttt{tests/} directory holding all
input/output pairs and a \texttt{test.sh} script that runs the agent's solution
and reports successes and failures on individual tests.\footnote{We put some
effort into building a robust test script that works even for a weak
model/agent. We set timeouts on each test, to help agents recover when they
write non-terminating programs. The test script takes care of recompilation when
needed, in case the agent forgets to compile. Our first version only printed
failure and used exit code 0 to indicate success. We found that \qwen and \gemma
had trouble with this format, so we switched to printing PASS and FAIL.} The
prompt explicitly states\footnote{See \cref{appendix:prompts} for the complete
prompt template.}:

\begin{quote}
\itshape
    To test your program, run exactly ./test.sh [...]
    These are the only tests I care about.
\end{quote}

We believe that this setup is a realistic way to study agent behavior: in
everyday use, programmers don't hide their tests from agents. Instead,
programmers direct their agents to keep working until all tests pass. It is this
type of task that MiniLCB simulates, and we get to observe how agents iterate on
their solutions when they have test feedback available. With this configuration,
there is a risk that the agent will cheat, but we only find cheating in 2 out of
all traces (\cref{appendix:cheating}).

% NOTE(arya): We do not stop on first failure.
There is also a risk that an agent will time out while running test cases, if it
writes an inefficient or non-terminating solution. To mitigate this, the
\texttt{test.sh} script imposes a 10-second timeout on each test case. An agent
run terminates when the agent submits its solution, exhausts its 40-turn budget,
or exceeds the 600\,s wall-clock limit.

%% ============================================================
\section{Results}
\label{sec:eval}

For a complex task, agents may engage in multiple turns of reasoning,
interleaving different kinds of behavior. Thus, when looking at agent behavior
across languages, we can look both at differences in the turns they take, and in
the content of those turns. We explore both aspects below.

\subsection{Token Cost and Accuracy by Language}

When comparing agent behavior across programming languages, the most basic
question is how many tokens are consumed to solve a problem, which we measure as the total output tokens generated over the trajectory (\cref{appendix:outputtoken}).

\paragraph{Median token cost and accuracy}

\begin{figure}
    \centering
    \includegraphics[width=0.95\linewidth]{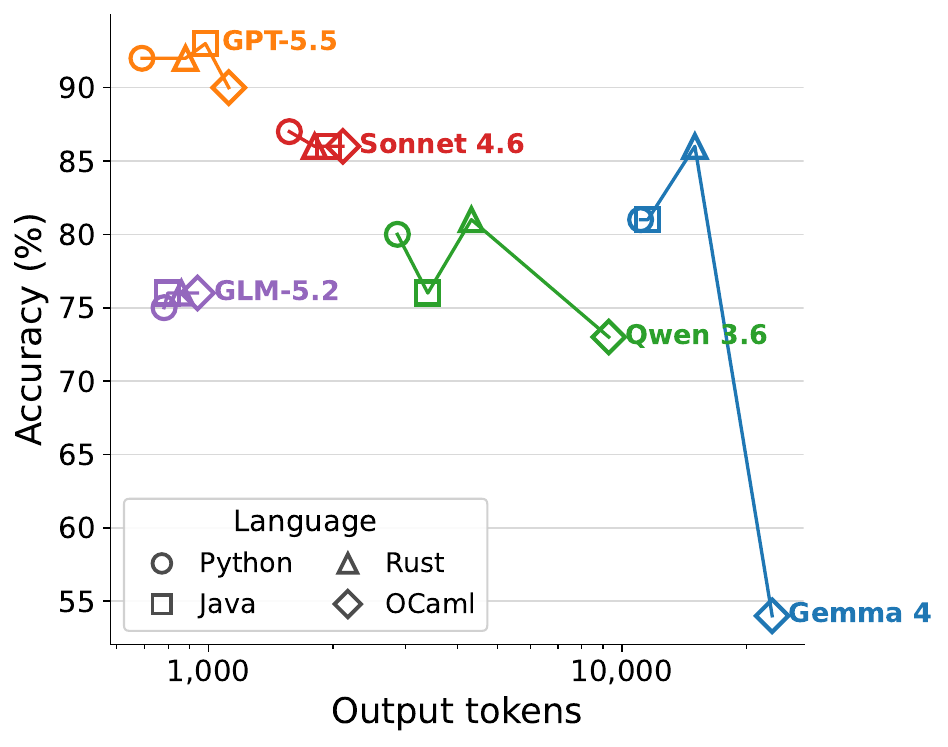}
    \caption{Accuracy versus median output token usage for each model and programming language. The output token axis is shown on a logarithmic scale.}
    \label{fig:acccuray-vs-output-tokens}
\end{figure}

In \Cref{fig:acccuray-vs-output-tokens} we report the median output token usage
and accuracy by language and model. For the models we explore, accuracy on
MiniLCB is not affected very much by language, with the notable exception that
\gemma performs much worse on OCaml. 
However, output tokens vary substantially by language. We report medians because
token costs are heavy-tailed: a few hard problems inflate the mean.
For all models, agents produce significantly more tokens when writing OCaml, and
the fewest tokens when writing Python.
We consider token usage across all traces, and not just the succeeding traces.
We believe this is realistic: if an agent runs for hours and fails, you don't get a refund.

% ── Table 2: Mixed-effects model ─────────────────────────────────────────────
\paragraph{Controlling for problem difficulty}
A limitation of the medians above is that they conflate programming-language
difficulty with problem difficulty, both of which affect output tokens.  To
isolate the effect of language, we fit 
\emph{mixed-effects models} for each LLM: a linear model that assigns a
baseline token cost to each problem, so the language coefficients reflect
cost differences \emph{on the same problem}.  
We treat language as a fixed effect with Python as the reference level, and the
choice of problem as a random effect. The intraclass correlation (ICC) measures
what fraction of token-cost variance is explained by problem identity; the
high values ($0.73$--$0.97$) confirm that problem difficulty dominates,
justifying why we control for it.

The mixed-effect model allows us to quantify the token penalty (or advantage)
for choosing a language other than Python, after controlling for problem
difficulty (\Cref{tab:mixed_effects,tab:mixed_effects_proprietary}). Across all
models, OCaml traces consume significantly more tokens than Python traces
($1.28$--$1.69\times$ vs.\ Python, $p<0.001$). The penalty for using Rust or
Java is model-dependent, but not significant for \gemma.

% TODO(arjun): Insert joke later as final sentence.
%Overall, these results strongly suggest that tokenmaxxers should not use Python, and of these languages, OCaml is best for tokenmaxxing.

\begin{table}[t]
\centering
\caption{Mixed-effects model results: multiplicative output token cost relative to Python, after controlling for problem difficulty. OCaml has a significant penalty with all models; Java and Rust effects are model-dependent.}
\label{tab:mixed_effects}
\small
\setlength{\tabcolsep}{3pt}
\begin{tabular}{lrll}
\toprule
Contrast & \llap{Token-cost ratio} & 95\% CI & Significance \\
\midrule
\multicolumn{4}{l}{\textbf{\gemma} \textit{(ICC 0.80)}} \\
  \textcolor{gray}{Java vs.\ Python}  & \textcolor{gray}{$0.85\times$} & \textcolor{gray}{$[0.70, 1.04]$} & \textcolor{gray}{$p=0.114$} \\
  \textcolor{gray}{Rust vs.\ Python}  & \textcolor{gray}{$1.07\times$} & \textcolor{gray}{$[0.88, 1.31]$} & \textcolor{gray}{$p=0.480$} \\
  OCaml vs.\ Python & $1.44\times$ & $[1.18, 1.75]$ & $p < 0.001$ \\
\midrule
\multicolumn{4}{l}{\textbf{\qwen} \textit{(ICC 0.89)}} \\
  Java vs.\ Python  & $1.21\times$ & $[1.03, 1.42]$ & $p < 0.05$ \\
  Rust vs.\ Python  & $1.20\times$ & $[1.02, 1.41]$ & $p < 0.05$ \\
  OCaml vs.\ Python & $1.69\times$ & $[1.44, 1.98]$ & $p < 0.001$ \\
\midrule
\multicolumn{4}{l}{\textbf{\glm} \textit{(ICC 0.73)}} \\
  Java vs.\ Python  & $1.34\times$ & $[1.09, 1.65]$ & $p < 0.01$ \\
  Rust vs.\ Python  & $1.57\times$ & $[1.28, 1.93]$ & $p < 0.001$ \\
  OCaml vs.\ Python & $1.54\times$ & $[1.25, 1.88]$ & $p < 0.001$ \\
\bottomrule
\end{tabular}
\end{table}

\begin{table}[t]
\centering
\caption{We run all intermediate solutions that the agent generates and report the number of tokens (turns) to the first solution that passes all tests, and the tokens (turns) spent after that solution.}
\label{tab:efficiency}
\small
\begin{tabular}{llrr}

\toprule
Model & Language &
 \makecell{Tokens (Turns) \\ to Solution} & \makecell{Tokens (Turns) \\ after Solution} \\
\midrule

% \toprule
% Model & Lang & $\text{Tokens}_{\rightarrow}$ & $\text{Turns}_{\rightarrow}$ & $\text{Tokens}_{+}$ & $\text{Turns}_{+}$ \\
% \midrule
\multirow{4}{*}{\gemma}
  & Python &  1820 (1) &  3628 (3) \\
  & Java   &  2498 (1) &  4628 (2) \\
  & Rust   &  2026 (1) &  7312 (3) \\
  & OCaml  &  2495 (2) &  3356 (3) \\
\midrule
\multirow{4}{*}{\qwen}
  & Python &  1756  (1) &   294  (3) \\
  & Java   &  2153  (1) &   272  (2) \\
  & Rust   &  2409  (3) &   370  (4) \\
  & OCaml  &  3330  (6) &   405  (5) \\
\midrule
\multirow{4}{*}{\glm}
  & Python &   758  (1) &   117  (2) \\
  & Java   &   730  (1) &    98  (1) \\
  & Rust   &  1574  (1) &   122  (2) \\
  & OCaml  &  1374  (3) &   109  (2) \\
\bottomrule
\end{tabular}
\end{table}
%% ------------------------------------------------------------
\subsection{Where Do All Those Tokens Go?}
\label{sec:analysis:formalism}
%% ------------------------------------------------------------

In the rest of this paper, we go beyond accuracy and try to shed light on what
these agents are doing with all of these tokens. We present the results for
open-weight models below; the corresponding results for proprietary models are
provided in \Cref{appendix:proprietary}.

% ── Table 3: Efficiency breakdown 
\subsubsection{Tokens after the first correct solution}
\label{tokens-after}

One surprisingly common behavior that we observe is that agents continue to
refine their solutions after they get it right. We study this behavior by splitting successful trajectories at the first correct
solution: the first intermediate solution that passes all tests, which the agent may not have tested itself.

\Cref{tab:efficiency} shows the median number of tokens that each model needed
to produce a working solution for each language and the median number of
additional tokens produced afterward. Our prompt directs the model to print a
sentinel message when it is done, which requires a few tokens. But, the smaller
models (\gemma and \qwen) produce hundreds or thousands more tokens after they
find their first working solution.

One can instead measure the number of tokens (turns) until the agent observes that all tests pass. This covers fewer solutions, because agents don't always run the provided test suite. With this measure, Gemma still runs for hundreds to thousands of tokens after observing that tests pass, while Qwen and GLM run for under a hundred (\cref{appendix:agent-observed}).

\begin{figure*}[t]
    \centering
    \begin{subfigure}[t]{0.47\textwidth}
        \centering
        \begin{tikzpicture}[
  transition node/.style={
    circle,
    minimum size=0.82cm,
    align=center,
    text=white,
    font=\tiny,
    inner sep=0pt,
    outer sep=2pt
  },
  terminal node/.style={
    rounded corners=2pt,
    minimum width=0.68cm,
    minimum height=0.52cm,
    align=center,
    text=white,
    font=\tiny,
    inner sep=1pt,
    outer sep=2pt,
    draw=slate edge
  },
  transition/.style={
    -{Stealth[length=1.9mm,width=1.4mm]},
    line cap=round,
    line join=round,
    shorten >=1pt,
    shorten <=1pt
  },
  edge count/.style={
    pos=0.78,
    fill=white,
    fill opacity=0.72,
    text opacity=1,
    inner sep=0.5pt,
    font=\tiny
  },
  edge count slate/.style={edge count, text=slate edge},
  edge count stuck/.style={edge count, text=stuck fill},
  edge count improved/.style={edge count, text=improved fill},
  edge count breakthrough/.style={edge count, text=breakthrough fill},
  edge count churn/.style={edge count, text=churn fill},
  edge count regressed/.style={edge count, text=regressed fill},
  edge count broke/.style={edge count, text=broke fill},
  edge count stay/.style={edge count, text=stay fill},
  x=1cm,
  y=1cm,
  node distance=1.05cm and 1.12cm
]
\definecolor{start fill}{RGB}{151,166,186}
\definecolor{slate edge}{RGB}{97,111,132}
\definecolor{improved fill}{RGB}{44,151,229}
\definecolor{stuck fill}{RGB}{255,156,11}
\definecolor{breakthrough fill}{RGB}{3,182,197}
\definecolor{churn fill}{RGB}{158,33,185}
\definecolor{regressed fill}{RGB}{246,65,57}
\definecolor{broke fill}{RGB}{142,0,0}
\definecolor{stay fill}{RGB}{76,174,78}
\definecolor{limit fill}{RGB}{151,166,186}

\node[terminal node, fill=start fill] (start) {start};
\node[transition node, fill=stuck fill, right=0.88cm of start] (stuck) {stuck};
\node[transition node, fill=improved fill, above=0.78cm of stuck] (improved) {improved};
\node[transition node, fill=breakthrough fill, below=0.78cm of stuck] (breakthrough) {break\\[-2pt]through};
\node[transition node, fill=churn fill, above right=0.34cm and 1.08cm of improved] (churn) {churn};
\node[transition node, fill=regressed fill, right=2.16cm of improved] (regressed) {regressed};
\node[transition node, fill=broke fill, right=2.16cm of stuck] (broke) {broke};
\node[transition node, fill=stay fill, right=2.16cm of breakthrough] (stay) {stay};
\node[terminal node, fill=limit fill, below right=0.58cm and 1.08cm of breakthrough] (limit) {limit};
\node[terminal node, fill=start fill, right=0.88cm of broke] (submit) {submit};

\path[transition, slate edge]
  (start) edge node[edge count slate, above, pos=0.5] {17} (stuck)
  (start) edge[bend left=10] node[edge count slate, above, pos=0.6] {10} (improved)
  (start) edge[bend right=8] node[edge count slate, below, pos=0.5] {73} (breakthrough);

\path[transition, stuck fill]
  (stuck) edge[out=140,in=220,looseness=6] node[edge count stuck, left] {52} (stuck)
  (stuck) edge[bend left=9] node[edge count stuck, left, pos=0.6] {17} (improved)
  (stuck) edge[bend right=10] node[edge count stuck, left, pos=0.6] {12} (breakthrough)
  (stuck) edge[bend left=22] node[edge count stuck, above] {2} (regressed)
  (stuck) edge[bend left=18] node[edge count stuck, above] {5} (broke)
  (stuck) edge[bend right=10] node[edge count stuck, below] {4} (limit)
  (stuck) edge[bend right=20] node[edge count stuck, above, pos=0.85] {7} (submit);

\path[transition, improved fill]
  (improved) edge[out=135,in=45,looseness=4] node[edge count improved, above] {4} (improved)
  (improved) edge[bend left=10] node[edge count improved, above, pos=0.6] {1} (churn)
  (improved) edge[bend left=6] node[edge count improved, above] {7} (regressed)
  (improved) edge[bend left=18] node[edge count improved, above, pos=0.78] {2} (broke)
  (improved) edge[bend right=3] node[edge count improved, below, pos=0.4] {4} (stuck)
  (improved) edge[bend right=34] node[edge count improved, left] {15} (breakthrough)
  (improved) edge[bend right=6] node[edge count improved, right, pos=0.9] {4} (limit)
  (improved) edge[bend left=15] node[edge count improved, above, pos=0.9] {1} (submit);

\path[transition, breakthrough fill]
  (breakthrough) edge[bend right=15] node[edge count breakthrough, above, pos=0.85] {3} (regressed)
  (breakthrough) edge[bend right=26] node[edge count breakthrough, above, pos=0.85] {1} (broke)
  (breakthrough) edge node[edge count breakthrough, above, pos=0.85] {54} (stay)
  (breakthrough) edge[bend right=4] node[edge count breakthrough, below] {40} (submit);

\path[transition, churn fill]
  (churn) edge[out=0,in=45,looseness=5] node[edge count churn, above] {1} (churn)
  (churn) edge[bend right=30] node[edge count churn, above, pos=0.78] {2} (broke)
  (churn) edge[bend right=2] node[edge count churn, right, pos=0.8] {1} (breakthrough);

\path[transition, regressed fill]
  (regressed) edge[out=35,in=145,looseness=5] node[edge count regressed, above] {1} (regressed)
  (regressed) edge[bend left=9] node[edge count regressed, below, pos=0.5] {2} (churn)
  (regressed) edge[bend right=18] node[edge count regressed, below, pos=0.77] {6} (stuck)
  (regressed) edge[bend left=18] node[edge count regressed, right] {1} (breakthrough)
  (regressed) edge[out=-50,in=15] node[edge count regressed, right] {1} (limit);

\path[transition, broke fill]
  (broke) edge[bend left=32] node[edge count broke, above, pos=0.85] {4} (improved)
  (broke) edge[bend left=18] node[edge count broke, above] {9} (stuck)
  (broke) edge[bend left=30] node[edge count broke, below, pos=0.78] {6} (breakthrough);

\path[transition, stay fill]
  (stay) edge[out=150,in=210,looseness=5] node[edge count stay, left] {205} (stay)
  (stay) edge[bend right=30] node[edge count stay, left] {2} (regressed)
  (stay) edge[bend right=18] node[edge count stay, right, pos=0.5] {9} (broke)
  (stay) edge[bend right=4] node[edge count stay, below] {34} (submit)
  (stay) edge[bend left=5] node[edge count stay, below, pos=0.5] {9} (limit);

\end{tikzpicture}
        \caption{\gemma Python}
    \end{subfigure}\hfill
    \begin{subfigure}[t]{0.47\textwidth}
        \centering
        \begin{tikzpicture}[
  transition node/.style={
    circle,
    minimum size=0.82cm,
    align=center,
    text=white,
    font=\tiny,
    inner sep=0pt,
    outer sep=2pt
  },
  terminal node/.style={
    rounded corners=2pt,
    minimum width=0.68cm,
    minimum height=0.52cm,
    align=center,
    text=white,
    font=\tiny,
    inner sep=1pt,
    outer sep=2pt,
    draw=slate edge
  },
  transition/.style={
    -{Stealth[length=1.9mm,width=1.4mm]},
    line cap=round,
    line join=round,
    shorten >=1pt,
    shorten <=1pt
  },
  edge count/.style={
    pos=0.78,
    fill=white,
    fill opacity=0.72,
    text opacity=1,
    inner sep=0.5pt,
    font=\tiny
  },
  edge count slate/.style={edge count, text=slate edge},
  edge count stuck/.style={edge count, text=stuck fill},
  edge count improved/.style={edge count, text=improved fill},
  edge count breakthrough/.style={edge count, text=breakthrough fill},
  edge count regressed/.style={edge count, text=regressed fill},
  edge count broke/.style={edge count, text=broke fill},
  edge count stay/.style={edge count, text=stay fill},
  node distance=1.05cm and 1.12cm
]
\definecolor{start fill}{RGB}{151,166,186}
\definecolor{slate edge}{RGB}{97,111,132}
\definecolor{improved fill}{RGB}{44,151,229}
\definecolor{stuck fill}{RGB}{255,156,11}
\definecolor{breakthrough fill}{RGB}{3,182,197}
\definecolor{regressed fill}{RGB}{246,65,57}
\definecolor{broke fill}{RGB}{142,0,0}
\definecolor{stay fill}{RGB}{76,174,78}
\definecolor{limit fill}{RGB}{151,166,186}

\node[terminal node, fill=start fill] (start) {start};
\node[transition node, fill=stuck fill, right=0.88cm of start] (stuck) {stuck};
\node[transition node, fill=improved fill, above=0.78cm of stuck] (improved) {improved};
\node[transition node, fill=breakthrough fill, below=0.78cm of stuck] (breakthrough) {break\\[-2pt]through};
\node[transition node, fill=regressed fill, right=2.16cm of improved] (regressed) {regressed};
\node[transition node, fill=broke fill, right=2.16cm of stuck] (broke) {broke};
\node[transition node, fill=stay fill, right=2.16cm of breakthrough] (stay) {stay};
\node[terminal node, fill=limit fill, below right=0.58cm and 1.08cm of breakthrough] (limit) {limit};
\node[terminal node, fill=start fill, right=0.88cm of broke] (submit) {submit};

\path[transition, slate edge]
  (start) edge[bend left=10] node[edge count slate, above, pos=0.5] {1} (improved)
  (start) edge node[edge count slate, above, pos=0.5] {78} (stuck)
  (start) edge[bend right=8] node[edge count slate, below, pos=0.5] {21} (breakthrough);

\path[transition, stuck fill]
  (stuck) edge[out=140,in=220,looseness=6] node[edge count stuck, left] {1196} (stuck)
  (stuck) edge[bend left=15] node[edge count stuck, left, pos=0.6] {26} (improved)
  (stuck) edge[bend right=10] node[edge count stuck, left,pos=0.50] {54} (breakthrough)
  (stuck) edge[bend left=18] node[edge count stuck, above] {14} (broke)
  (stuck) edge[bend right=20] node[edge count stuck, above] {42} (limit);

\path[transition, improved fill]
  (improved) edge[bend right=3] node[edge count improved, left, pos=0.6] {20} (stuck)
  (improved) edge[bend right=30] node[edge count improved, right] {1} (breakthrough)
  (improved) edge[bend left=8] node[edge count improved, above,pos=0.8] {13} (broke)
  (improved) edge[bend right=5] node[edge count improved, below, pos=0.8] {2} (limit);

\path[transition, breakthrough fill]
  (breakthrough) edge[bend left=12] node[edge count breakthrough, above, pos=0.9] {2} (regressed)
  (breakthrough) edge[bend right=5] node[edge count breakthrough, above] {21} (broke)
  (breakthrough) edge node[edge count breakthrough, above, pos=0.85] {28} (stay)
  (breakthrough) edge[bend right=10] node[edge count breakthrough, above, pos=0.9] {35} (submit)
  (breakthrough) edge[bend right=6] node[edge count breakthrough, below, pos=0.6] {3} (limit);

\path[transition, regressed fill]
  (regressed) edge[bend right=16] node[edge count regressed, below] {1} (stuck)
  (regressed) edge[bend left=20] node[edge count regressed, right, pos=0.5] {2} (broke);

\path[transition, broke fill]
  (broke) edge[bend left=22] node[edge count broke, above,] {9} (improved)
  (broke) edge[bend left=12] node[edge count broke, above] {37} (stuck)
  (broke) edge[bend left=10] node[edge count broke, below] {13} (breakthrough)
  (broke) edge[bend left=12] node[edge count broke, right] {1} (limit);

\path[transition, stay fill]
  (stay) edge[out=150,in=210,looseness=5] node[edge count stay, left] {56} (stay)
  (stay) edge[bend right=30] node[edge count stay, right] {1} (regressed)
  (stay) edge[bend right=14] node[edge count stay, right, pos=0.5] {10} (broke)
  (stay) edge[bend right=4] node[edge count stay, below] {14} (submit)
  (stay) edge[bend left=5] node[edge count stay, below, pos=0.5] {3} (limit);
\end{tikzpicture}
        \caption{\gemma OCaml}
    \end{subfigure}

    \vspace{0.5em}

    \begin{subfigure}[t]{0.47\textwidth}
        \centering
        \begin{tikzpicture}[
  transition node/.style={
    circle,
    minimum size=0.82cm,
    align=center,
    text=white,
    font=\tiny,
    inner sep=0pt,
    outer sep=2pt
  },
  terminal node/.style={
    rounded corners=2pt,
    minimum width=0.68cm,
    minimum height=0.52cm,
    align=center,
    text=white,
    font=\tiny,
    inner sep=1pt,
    outer sep=2pt,
    draw=slate edge
  },
  transition/.style={
    -{Stealth[length=1.9mm,width=1.4mm]},
    line cap=round,
    line join=round,
    shorten >=1pt,
    shorten <=1pt
  },
  edge count/.style={
    pos=0.78,
    fill=white,
    fill opacity=0.72,
    text opacity=1,
    inner sep=0.5pt,
    font=\tiny
  },
  edge count slate/.style={edge count, text=slate edge},
  edge count breakthrough/.style={edge count, text=breakthrough fill},
  node distance=1.05cm and 1.12cm
]
\definecolor{start fill}{RGB}{151,166,186}
\definecolor{slate edge}{RGB}{97,111,132}
\definecolor{breakthrough fill}{RGB}{3,182,197}
\definecolor{limit fill}{RGB}{151,166,186}

\node[terminal node, fill=start fill] (start) {start};
\node[transition node, fill=breakthrough fill, right=0.88cm of start] (breakthrough) {break\\[-2pt]through};
\node[terminal node, fill=start fill, right=1.20cm of breakthrough] (submit) {submit};
\node[terminal node, fill=limit fill, right=0.72cm of submit] (limit) {limit};

\begin{pgfinterruptboundingbox}
\path[transition, slate edge]
  (start) edge node[edge count slate, above, pos=0.5] {75} (breakthrough);

\path[transition, breakthrough fill]
  (breakthrough) edge node[edge count breakthrough, above] {74} (submit)
  (breakthrough) edge[bend left=22] node[edge count breakthrough, above] {1} (limit);

\end{pgfinterruptboundingbox}
\end{tikzpicture}
        \caption{\glm Python}
    \end{subfigure}\hfill
    \begin{subfigure}[t]{0.47\textwidth}
        \centering
        \begin{tikzpicture}[
  transition node/.style={
    circle,
    minimum size=0.82cm,
    align=center,
    text=white,
    font=\tiny,
    inner sep=0pt,
    outer sep=2pt
  },
  terminal node/.style={
    rounded corners=2pt,
    minimum width=0.68cm,
    minimum height=0.52cm,
    align=center,
    text=white,
    font=\tiny,
    inner sep=1pt,
    outer sep=2pt,
    draw=slate edge
  },
  transition/.style={
    -{Stealth[length=1.9mm,width=1.4mm]},
    line cap=round,
    line join=round,
    shorten >=1pt,
    shorten <=1pt
  },
  edge count/.style={
    pos=0.78,
    fill=white,
    fill opacity=0.72,
    text opacity=1,
    inner sep=0.5pt,
    font=\tiny
  },
  edge count slate/.style={edge count, text=slate edge},
  edge count stuck/.style={edge count, text=stuck fill},
  edge count improved/.style={edge count, text=improved fill},
  edge count breakthrough/.style={edge count, text=breakthrough fill},
  edge count stay/.style={edge count, text=stay fill},
  node distance=1.05cm and 1.12cm
]
\definecolor{start fill}{RGB}{151,166,186}
\definecolor{slate edge}{RGB}{97,111,132}
\definecolor{improved fill}{RGB}{44,151,229}
\definecolor{stuck fill}{RGB}{255,156,11}
\definecolor{breakthrough fill}{RGB}{3,182,197}
\definecolor{stay fill}{RGB}{76,174,78}
\definecolor{limit fill}{RGB}{151,166,186}

\node[terminal node, fill=start fill] (start) {start};
\node[transition node, fill=stuck fill, right=0.88cm of start] (stuck) {stuck};
\node[transition node, fill=improved fill, right=0.88cm of stuck] (improved) {improved};
\node[transition node, fill=breakthrough fill, below=0.78cm of stuck] (breakthrough) {break\\[-2pt]through};
\node[terminal node, fill=start fill, right=1.20cm of breakthrough] (submit) {submit};
\node[terminal node, fill=limit fill, right=0.72cm of submit] (limit) {limit};

\begin{pgfinterruptboundingbox}
\path[transition, slate edge]
  (start) edge[bend left=18] node[edge count slate, above, pos=0.5] {2} (improved)
  (start) edge node[edge count slate, above, pos=0.5] {23} (stuck)
  (start) edge[bend right=8] node[edge count slate, below, pos=0.5] {52} (breakthrough);

\path[transition, stuck fill]
  (stuck) edge[out=140,in=220,looseness=6] node[edge count stuck, left] {7} (stuck)
  (stuck) edge[bend right=10] node[edge count stuck, left, pos=0.5] {22} (breakthrough)
  (stuck) edge[bend left=12] node[edge count stuck, above, pos=0.7] {1} (limit);

\path[transition, improved fill]
  (improved) edge[bend left=12] node[edge count improved, right] {2} (breakthrough)
  ;

\path[transition, breakthrough fill]
  (breakthrough) edge node[edge count breakthrough, above] {76} (submit);

\end{pgfinterruptboundingbox}
\end{tikzpicture}
        \caption{\glm OCaml}
    \end{subfigure}
    \caption{We test every solution that the agent writes to the solution file, which allows us to abstract each trajectory into a list of \emph{solution snapshots} with test outcomes. We then classify each consecutive pair of test outcomes based on how outcomes change from the previous snapshot to the current snapshot. We describe the outcome labels in detail in ~\Cref{graph-analysis}. The \emph{breakthrough} and \emph{stay} nodes are those where all tests pass in the snapshot, and all other colored nodes are cases where some tests fail. We consolidate trajectories for model-language pairs into individual graphs to see how models behave on each language.}
    \label{fig:transition_grid}
\end{figure*}
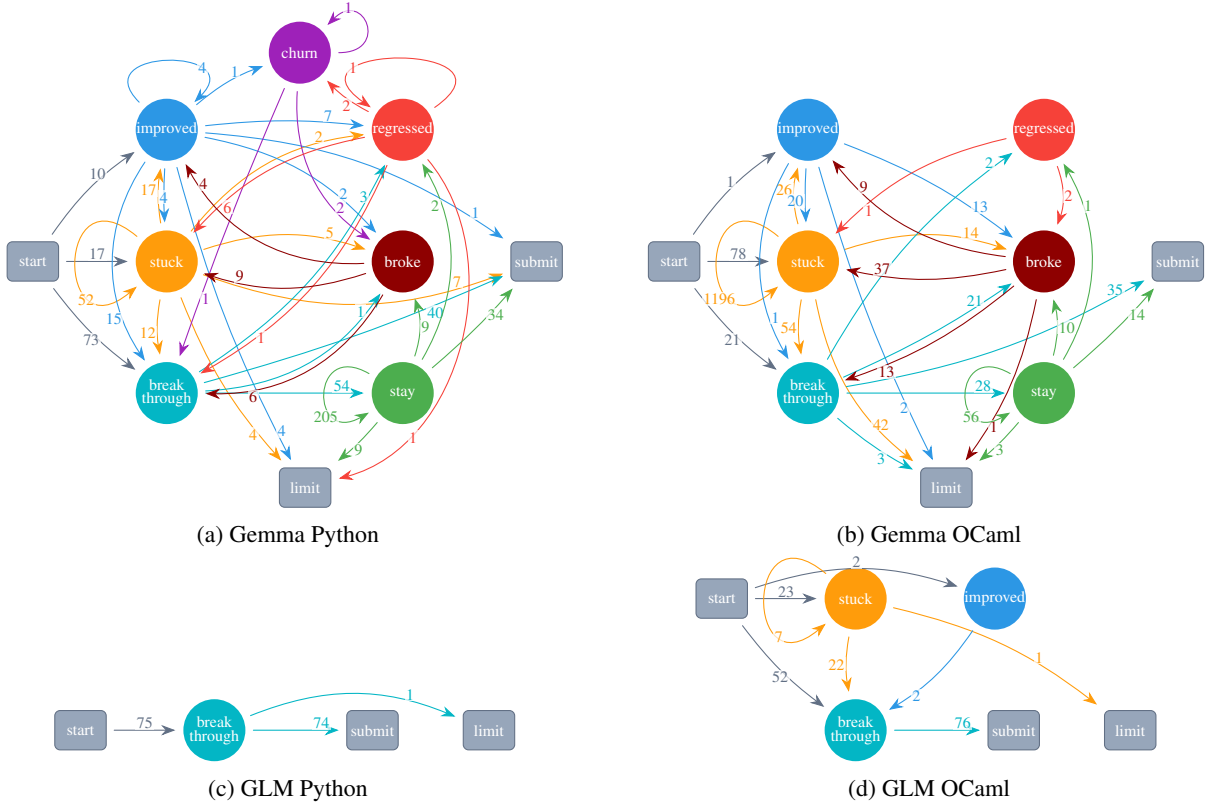

\subsubsection{Progression on provided unit tests}
\label{graph-analysis}

We now analyze trajectories by running the test suite on
every intermediate solution that the agent writes to disk and record the list of
test outcomes.\footnote{The solution is the full on-disk state needed to reconstruct the intermediate point in the agent's solution, which we can read off the saved trajectory.} This allows us to abstract each trajectory to a
list of \emph{solution snapshots}, $s_0, s_1, \ldots, s_n$, where each $s_i$ is
a list of test outcomes. We then classify each  consecutive pair $(s_{i-1},
s_i)$ based on how the test outcomes change from  $s_{i-1}$ to $s_i$:
\begin{itemize}[nosep,leftmargin=1.5em]
  \item \emph{breakthrough}: some tests were failing and all now pass;
  \item \emph{stay}: all tests passed before and after, i.e., the model rewrote a correct solution;
  \item \emph{improved}: some failing tests now pass (not all);
  \item \emph{regressed}: some passing tests now fail (not all);
  \item \emph{churn}: some passing tests now fail, and failing tests now pass (improvement and regression);
  \item \emph{broke}: some tests were passing, but now all fail, e.g., a compile error;
  \item \emph{stuck}: some tests fail, and no change in test outcomes;
\end{itemize}

We build a graph where nodes represent these outcome changes, as well as special nodes for the start state, a submission (which may be wrong!), or reaching a resource limit.

% Both models enter \textit{stuck}
% more frequently on the lower-resourced OCaml, with a subset of problems looping on
% \textit{stuck} before recovering.  Gemma additionally exhibits heavy
% post-success looping on \textit{stay}, and particularly on Python, reflecting continued
% cosmetic and optimization edits after already reaching a correct solution. 

In \cref{fig:transition_grid}, we show the graphs for \gemma and \glm on Python
and OCaml. (The corresponding \gpt graphs appear in
\cref{appendix:proprietary}.) These graphs allow us to quickly discover
interesting behavior in the models. For example, \gemma has exactly one
\textit{breakthrough}$\to$\textit{broke} edge across all Python problems in a
trajectory that has this shape: \\
\begin{tikzpicture}[
  trajectory node/.style={
    circle,
    minimum size=0.72cm,
    text width=0.64cm,
    align=center,
    text=white,
    font=\tiny,
    inner sep=0pt,
    outer sep=1pt
  },
  trajectory terminal/.style={
    rounded corners=1.5pt,
    minimum width=0.42cm,
    minimum height=0.38cm,
    align=center,
    text=white,
    font=\tiny,
    inner sep=1pt,
    outer sep=1pt,
    draw=slate edge
  },
  trajectory arrow/.style={
    -{Stealth[length=1.4mm,width=1.0mm]},
    slate edge,
    line cap=round,
    line join=round,
    shorten >=0.8pt,
    shorten <=0.8pt
  },
  loop count/.style={
    pos=0.5,
    font=\tiny,
    text=black,
    inner sep=0pt
  },
  node distance=0.22cm and 0.34cm
]
\definecolor{start fill}{RGB}{151,166,186}
\definecolor{slate edge}{RGB}{97,111,132}
\definecolor{improved fill}{RGB}{44,151,229}
\definecolor{stuck fill}{RGB}{255,156,11}
\definecolor{breakthrough fill}{RGB}{3,182,197}
\definecolor{regressed fill}{RGB}{246,65,57}
\definecolor{broke fill}{RGB}{142,0,0}

\node[trajectory node, fill=breakthrough fill] (breakthrough1) {break\\[-2pt]through};
\node[trajectory node, fill=broke fill, right=of breakthrough1] (broke) {broke};
\node[trajectory node, fill=stuck fill, right=of broke] (stuck2) {stuck};
\node[trajectory node, fill=breakthrough fill, right=of stuck2] (breakthrough2) {break\\[-2pt]through};
\node[trajectory node, fill=regressed fill, right=of breakthrough2] (regressed) {regressed};
\node[trajectory node, fill=stuck fill, right=of regressed] (stuck3) {stuck};
\node[trajectory terminal, fill=start fill, right=of stuck3] (submit) {submit};

\path[trajectory arrow]
  (breakthrough1) edge (broke)
  (broke) edge (stuck2)
  (stuck2) edge (breakthrough2)
  (breakthrough2) edge (regressed)
  (regressed) edge (stuck3)
  (stuck3) edge (submit);

\path[trajectory arrow]
  (stuck2) edge[out=120,in=60,looseness=4] node[loop count, above] {1x} (stuck2);
\end{tikzpicture}\\
Examining it reveals performance anxiety: despite the breakthrough, the model's reasoning text states:
{\setlength{\leftmargini}{0.5em}%
\begin{quote}
\itshape seems to be correct.  However, it might be too slow for the given constraints---let's optimize
\end{quote}}
Unfortunately, the optimization fails all tests. It corrects it (the second breakthrough), then regresses, and finally submits broken code. 

\begin{table*}[t]
\centering
\footnotesize
\setlength{\tabcolsep}{1.8pt}
\renewcommand{\arraystretch}{1.12}
\newcommand{\gaphead}[1]{\tiny\bfseries\makecell[cc]{#1}}
\begin{tabular}{@{}>{\raggedright\arraybackslash}m{0.09\textwidth}>{\raggedright\arraybackslash}m{0.048\textwidth}*{12}{>{\centering\arraybackslash}m{0.057\textwidth}}@{}}
\toprule
\textbf{Model} & \textbf{Lang} &
\gaphead{Implement} &
\gaphead{Explore} &
\gaphead{Rewrite} &
\gaphead{Execute\\plan} &
\gaphead{Fix\\bug} &
\gaphead{Debug\\failure} &
\gaphead{Infra\\confusion} &
\gaphead{Revert} &
\gaphead{Verify} &
\gaphead{Edge\\case} &
\gaphead{Optimize\\perf.} &
\gaphead{Cosmetic\\refactor} \\
\midrule
  \textbf{\gemma} & java & 31\% & 9\% & 7\% & <1\% & 7\% & 4\% & 1\% & <1\% & 27\% & 2\% & 4\% & 7\% \\
   & ocaml & 17\% & 6\% & 2\% & 1\% & 15\% & 15\% & 5\% & 24\% & 7\% & 1\% & <1\% & 5\% \\
   & python & 26\% & 8\% & 4\% & 2\% & 7\% & 3\% & 1\% & 3\% & 19\% & 1\% & 10\% & 16\% \\
   & rust & 28\% & 10\% & 5\% & 1\% & 6\% & 2\% & 1\% & 2\% & 24\% & 3\% & 4\% & 15\% \\
\midrule
  \textbf{\qwen} & java & 36\% & 5\% & 8\% & <1\% & 15\% & 7\% & 1\% & <1\% & 27\% & -- & <1\% & <1\% \\
   & ocaml & 19\% & 5\% & 5\% & 1\% & 36\% & 8\% & 12\% & -- & 12\% & -- & -- & 2\% \\
   & python & 35\% & 4\% & 8\% & <1\% & 14\% & 4\% & -- & -- & 29\% & -- & 4\% & 1\% \\
   & rust & 36\% & 4\% & 8\% & <1\% & 15\% & 3\% & 1\% & -- & 29\% & 1\% & <1\% & 2\% \\
\midrule
  \textbf{\glm} & java & 50\% & 1\% & -- & -- & -- & -- & -- & -- & 49\% & -- & -- & -- \\
   & ocaml & 42\% & -- & -- & -- & 13\% & 1\% & 3\% & -- & 42\% & -- & -- & 1\% \\
   & python & 50\% & -- & -- & -- & -- & -- & -- & -- & 50\% & -- & -- & -- \\
   & rust & 48\% & 1\% & -- & -- & 3\% & -- & -- & -- & 48\% & -- & -- & -- \\
\bottomrule
\end{tabular}
\caption{Percentage of \emph{spans} attributed to each label, showing how frequent each behavior is.  \glm is mostly \emph{implement}\,+\,\emph{verify}; debugging labels concentrate
  in OCaml; post-success labels are Gemma-dominated.}
\label{tab:gap-dist}
\end{table*}

\subsubsection{Text analysis of snapshot spans}
\label{span-labels}

The test outcomes of the solution snapshots show stages in the agent's progress, but they do not reveal what kinds of actions the agent takes during these stages. A natural unit for analysis is the turn, but turns are not semantically coherent: an agent
often needs several turns to accomplish one action: running a test,
reading the output, and correcting the code are three turns that make one logical bugfix action. We therefore analyze at the level of \emph{snapshot
spans}.  A span consists of everything the agent does between two
consecutive solution snapshots: the resulting solution snapshot, the unified
diff to the previous snapshot, the change in test outcomes, and the agent's
reasoning, bash command tool calls, and outputs up to and including the ending write. 

We
develop a taxonomy of twelve span labels by hand-examining a sample
of trajectories and iterating until the labels were mutually exclusive and
covered the full range of behaviors. We then label all spans by prompting an LLM judge (GPT-5-Mini) with the taxonomy and the span content and manually review all LLM-assigned labels:
\begin{itemize}[nosep,leftmargin=1.5em]
  \item \emph{explore}: working out a tentative approach --- stubs, comments, or exploratory scripts;
  \item \emph{implement}: commits real algorithmic logic, building on the current approach;
  \item \emph{rewrite}: abandons the current approach and rewrites with a new algorithm;
  \item \emph{execute plan}: fills in stubs or \texttt{\#TODO} placeholders from the prior snapshot with real logic;
  \item \emph{fix bug}: a focused patch with a named cause;
  \item \emph{debug failure}: tries to fix a bug without a named cause (e.g., speculative tweaks, debug prints);
  \item \emph{infra confusion}: struggling with scaffolding rather than the algorithm: wrong file conventions, misread test output, or build plumbing;
  \item \emph{revert}: returns to a previously seen solution, discarding recent changes;
  \item \emph{verify}: checks the solution's correctness;
  \item \emph{edge case}:  considers possible edge cases;
  \item \emph{optimize perf}: improves time/memory;
  \item \emph{cosmetic refactor}: renaming, reformatting, or commenting with no behavioral change.
\end{itemize}

For example, the following three-turn span is labeled \emph{fix bug}:
\begin{transcript}[traj:fix-bug-gap]
\assistant Let me test this optimized solution with the sample cases first, then run the full test suite.

\assistant The output is wrong. Let me debug...

\assistant I see the bug ... Let me fix:
\end{transcript}\vspace{0.5em}

The next two-turn span is an \emph{implement}, where \qwen{} first
solves the problem in Python and then translates to OCaml:
\begin{transcript}
\assistant The v3 formula passes all 31 tests. Let me verify it more carefully and then implement it in OCaml:
\begin{lstlisting}
cat << 'PYEOF' > analyze7.py
...
\end{lstlisting}

\assistant Now let me implement the solution in OCaml:
\end{transcript}\vspace{0.5em}

For every model--language pair, we report the breakdown of span labels---the share of spans attributed to each label, showing how frequent each behavior is---in
\Cref{tab:gap-dist}. \Cref{tab:gap-dist-tokens} gives the token-weighted version, which weights each span by its output tokens to show how much each behavior accounts for the total token expenditure. Across models, error-recovery labels---\emph{fix bug},
\emph{debug failure}, \emph{infra confusion}, and \emph{revert}---are
disproportionately associated with OCaml, suggesting that solving OCaml problems
requires more iterative correction. Post-success labels (\emph{optimize perf},
\emph{cosmetic refactor}) are most prevalent for \gemma, consistent with its
large post-solution token costs in \Cref{tab:efficiency}.

\subsection{How Do Agents Get (Un)stuck?}
\label{sec:analysis:stuck}
We can use the span labels (\cref{span-labels}) to ask how agents get unstuck on problems, using \gemma on OCaml as an example. This model-language pair has 1{,}196 self-loops on the \emph{stuck} state, dominated by \emph{revert} (31\%): the agent rewrites a byte-identical solution that it has already tried, oscillated between 2-3 broken attempts.
The agent is fighting OCaml syntax, not the algorithm:  92\% of failed snapshots are compile-time errors and 8\% are
wrong-answer failures.  The model rarely gets far enough to learn whether
its algorithm is correct.  Escape, when it comes, is a \emph{language
learning event}: \emph{fix bug} dominates breakthrough at 60\%
(Table~\ref{tab:stuck-breakthrough-labels}), the moment the agent finally
names a concrete OCaml-specific error and the unchanged algorithm
immediately passes.  Python breakthroughs, by contrast, are spread across
\emph{implement}, \emph{execute plan}, and
\emph{optimize perf}---there is no single bottleneck to escape.

On task \texttt{abc311\_d}, for instance, \gemma loops on stuck 14 times before escaping: \\
\begin{tikzpicture}[
  trajectory node/.style={
    circle,
    minimum size=0.72cm,
    text width=0.64cm,
    align=center,
    text=white,
    font=\tiny,
    inner sep=0pt,
    outer sep=1pt
  },
  trajectory terminal/.style={
    rounded corners=1.5pt,
    minimum width=0.42cm,
    minimum height=0.38cm,
    align=center,
    text=white,
    font=\tiny,
    inner sep=1pt,
    outer sep=1pt,
    draw=slate edge
  },
  trajectory arrow/.style={
    -{Stealth[length=1.4mm,width=1.0mm]},
    line cap=round,
    line join=round,
    shorten >=0.8pt,
    shorten <=0.8pt
  },
  loop count/.style={
    pos=0.5,
    font=\tiny,
    text=black,
    inner sep=0pt
  },
  node distance=0.22cm and 0.34cm
]
\definecolor{start fill}{RGB}{151,166,186}
\definecolor{slate edge}{RGB}{97,111,132}
\definecolor{improved fill}{RGB}{44,151,229}
\definecolor{stuck fill}{RGB}{255,156,11}
\definecolor{breakthrough fill}{RGB}{3,182,197}
\definecolor{regressed fill}{RGB}{246,65,57}
\definecolor{broke fill}{RGB}{142,0,0}
\definecolor{stay fill}{RGB}{76,174,78}

\node[trajectory terminal, fill=start fill] (start) {$\blacktriangleright$};
\node[trajectory node, fill=stuck fill, right=of start] (stuck) {stuck};
\node[trajectory node, fill=breakthrough fill, right=of stuck] (breakthrough) {break\\[-2pt]through};
\node[trajectory node, fill=stay fill, right=of breakthrough] (stay) {stay};
\node[trajectory terminal, fill=start fill, right=of stay] (submit) {submit};

\path[trajectory arrow, stuck fill]
  (stuck) edge[out=120,in=60,looseness=4] node[loop count, above] {14x} (stuck);

\path[trajectory arrow, stuck fill]
  (start) edge (stuck);

\path[trajectory arrow, stuck fill]
  (stuck) edge (breakthrough);

\path[trajectory arrow, breakthrough fill]
  (breakthrough) edge (stay);

\path[trajectory arrow, slate edge]
  (stay) edge (submit);
\end{tikzpicture}
\\
\gemma keeps the same algorithm throughout but cannot get it to compile,
cycling on the same OCaml type errors. Turns 8--14 oscillate between two earlier attempts,
re-emitting byte-for-byte identical code, until it finally names the error
and the unchanged algorithm compiles and passes all tests.

This section uses raw counts of self-loops, which can in principle be dominated by a few long trajectories. \Cref{appendix:per-problem} shows that the same phenomenon holds when each problem is weighted equally, and this applies for the next two sections as well.

\subsection{Cosmetic Refactoring}
\label{sec:analysis:stay}

As a second case study, we examine how \gemma{} revises perfect solutions (\textit{stay} loops).
Figure~\ref{fig:transition_grid} shows 205 stay self-loops in Python,
the most of any language.
These loops are unnecessary polishing: \emph{cosmetic refactor} and
\emph{optimize perf} account for half the edges into \textsc{stay}
(Table~\ref{tab:stay-labels}). For example, \gemma solves \texttt{arc194\_e} in
Python on Turn 2, but then second-guesses its solution and spends the remaining
turns on polish before a timeout:

\begin{tikzpicture}[
  trajectory node/.style={
    circle,
    minimum size=0.72cm,
    text width=0.64cm,
    align=center,
    text=white,
    font=\tiny,
    inner sep=0pt,
    outer sep=1pt
  },
  trajectory terminal/.style={
    rounded corners=1.5pt,
    minimum width=0.42cm,
    minimum height=0.38cm,
    align=center,
    text=white,
    font=\tiny,
    inner sep=1pt,
    outer sep=1pt,
    draw=slate edge
  },
  trajectory arrow/.style={
    -{Stealth[length=1.4mm,width=1.0mm]},
    line cap=round,
    line join=round,
    shorten >=0.8pt,
    shorten <=0.8pt
  },
  loop count/.style={
    pos=0.5,
    font=\tiny,
    text=black,
    inner sep=0pt
  },
  node distance=0.22cm and 0.34cm
]
\definecolor{start fill}{RGB}{151,166,186}
\definecolor{slate edge}{RGB}{97,111,132}
\definecolor{improved fill}{RGB}{44,151,229}
\definecolor{stuck fill}{RGB}{255,156,11}
\definecolor{breakthrough fill}{RGB}{3,182,197}
\definecolor{regressed fill}{RGB}{246,65,57}
\definecolor{broke fill}{RGB}{142,0,0}
\definecolor{stay fill}{RGB}{76,174,78}

\node[trajectory terminal, fill=start fill] (start) {$\blacktriangleright$};
\node[trajectory node, fill=improved fill, right=of start] (improved) {improved};
\node[trajectory node, fill=breakthrough fill, right=of improved] (breakthrough) {break\\[-2pt]through};
\node[trajectory node, fill=stay fill, right=of breakthrough] (stay) {stay};
\node[trajectory terminal, fill=start fill, right=of stay] (limit) {limit};

\path[trajectory arrow, stay fill]
  (stay) edge[out=120,in=60,looseness=4] node[loop count, above] {9x} (stay);

\path[trajectory arrow, improved fill]
  (start) edge (improved);

\path[trajectory arrow, breakthrough fill]
  (improved) edge (breakthrough);

\path[trajectory arrow, stay fill]
  (breakthrough) edge (stay);

\path[trajectory arrow, slate edge]
  (stay) edge (limit);
\end{tikzpicture}

\begin{transcript}
\assistant \textbf{Turn 3/12.} The logic was not quite right. I need to rewrite.
\ldots
\assistant \textbf{Turn 9/12.} My current code seems to be working on these
samples. Is there any other condition?
\ldots
\assistant \textbf{Turn 12/12.} Okay, the current solution seems to be working for the sample cases. \emph{``Wait, the input
could be quite large''} \ldots

\end{transcript}\vspace{0.5em}
Turn~3 reformulates the correct solution, Turn~9 adds only inline comments, and
Turn~12 changes the recursion depth, even though the solution is not recursive.
It hits the timeout and never submits.

\subsection{Text-Level Analysis}
The case studies above analyzed the \emph{structure} of trajectories---which nodes agents visit and how they move between them. We now move beyond the graph to analyze the \emph{content}: the text agents produce as they work through a problem.

\paragraph{Planning in code comments}
Models sometimes solve problems in comments before writing code.
E.g., on \texttt{abc359\_d}, \gemma spends two turns deriving a DP solution in comments, before translating it into working code. On \texttt{arc194\_a}, \qwen writes a 35-line Java solution with 3{,}439 lines of comments.

% Here is an example line from the file:

% \begin{lstlisting}[style=codeblock]
% print("Insight found -- need to track f[k] with M maintained")
% \end{lstlisting}

\paragraph{Solve in Python, then translate.}
Models sometimes solve problems in Python before translating the solution to an unfamiliar target language. E.g., to solve
\texttt{abc359\_c} in OCaml, \qwen first takes 12 turns to write a Python solution, including a translated test harness. Similarly, to solve \texttt{arc194\_e} in Rust, \glm
spends 11 turns in Python testing mathematical invariants against a brute-force
checker before writing a Rust solution that works immediately. It then returns to Python to stress-test the solution on
large inputs before submitting.

\paragraph{Second-guessing in reasoning.}
Reasoning traces show that models frequently reconsider their
approach. \gemma and \qwen reason very reflectively, frequently using words such
as \emph{wait} and \emph{actually} (7.70 and 7.72 occurrences per turn,
respectively), alongside frequent mentions of correctness (3.89 and 1.70) and
efficiency (2.02 and 1.65). \glm shows the same tendency more moderately (4.27
second-guessing, 1.65 correctness, 0.87 efficiency mentions per turn). These
keywords appear in both affirmative and negative contexts (e.g.,
\emph{``correct''} and \emph{``incorrect''}), suggesting that the models
repeatedly reevaluate whether their current solution is correct. The trajectory
example in \Cref{sec:analysis:stay} illustrates this behavior: after producing a
solution that retrospectively passes all tests, \gemma remarks, \emph{``Wait,
the input could be quite large,''} and repeatedly rewrites the solution.

\paragraph{Testing and self-examination.}
\gemma and \qwen
often rewrite solutions without running the provided tests: over 70\% of
their snapshot spans into \textsc{stay} are bare solution writes with no
preceding test execution. Instead, they frequently challenge their solutions by
replaying the sample inputs or constructing their own test cases. 

For example, on \texttt{arc194\_e}, \gemma repeatedly revises the solution using tests that it writes itself. However, it continues to do so even after it produces a solution that passes all provided tests---unfortunately, it never runs the provided tests on this problem despite  directions to do so. This is in contrast to \glm, which uses the provided tests to stop much earlier.

%% ============================================================
\section{Conclusion}
\label{sec:conclusion}
%% ============================================================

We examine how the choice of programming language affects the behavior of coding
agents. Our most basic finding is that token consumption can vary significantly
by programming language, and agents can spend far more tokens on a
lower-resource language than higher-resource languages.

What is more revealing is what the agent is doing with its tokens on each
language. We investigate this by abstracting agent trajectories into
trajectories of intermediate solutions and their test results. This analysis
exposes a range of token waste patterns---agents polishing solutions after all
tests pass, and fighting a low-resource language's syntax rather than improving
the solution's algorithm---and also more deliberate strategies, such as planning
in code comments and prototyping in Python before translating to an unfamiliar
target. We hope these findings point the way toward designing agents that work
more efficiently across all programming languages.

\label{limitcheck}
\makeatletter
\ifnum\getpagerefnumber{limitcheck}>8 % <-- 8 page for long paper
  \begin{center}
    \color{red}\Large\bfseries
    [ACL Has 8 Page Limit]
  \end{center}
\fi
\makeatother

\section*{Limitations}
Our study covers four programming languages and five models; the behaviors we identify may not generalize to other models, and the language-specific effects we observe may change as future models improve pretraining coverage of lower-resource languages. MiniLCB consists of self-contained competitive-programming problems with visible test cases, solved using a single agent scaffold (mini-swe-agent) with a fixed prompt, turn limit, and timeout. Our findings may therefore not generalize to repository-scale software engineering tasks, or to other agent harnesses. We deliberately choose a setting where an agent should be best positioned to check its own work---all tests are visible, and a mechanism to run them is provided---yet we still observe agents spending turns on unrelated work instead of checking their solution, and continuing to revise a solution after already observing it pass. We expect this inefficiency to be more pronounced, not less, in less constrained real development settings.

\section*{Ethical Considerations}

Our paper title should not be read literally. Our intent is to understand why agents use more tokens in low-resource programming languages so that more efficient multilingual agents can be built. We discourage using these findings to deliberately increase token consumption.

\ifanon\else
\section*{Acknowledgments}
This work is partially supported by the National Science Foundation (SES-2326173), an Oracle Collaborative Research Award, an OpenAI credit award from the NAIRR Pilot (NAIRR250283), and by the NVIDIA Academic Grant Program using a DGX Spark.
\fi

%% ============================================================
%%  Bibliography
%%  Put your .bib file name(s) below (without the .bib extension).
%% ============================================================
\bibliography{arya,main}

\clearpage

\appendix

\section{Hyperparameters and Configurations}
\label{appendix:hyperparams}

% \ag{Maybe we should put appendices before the references, and fit them in the 10 page limit. FP8 for Gemma 4. Max length. Generation hyperparams for all.

% 10MB output cap. Boa specifications. Output tokens/second. 

% Hmm too much detail. Definitely do not include the actual serving command or the machine name. Total number of GPUs doesn't matter. OS doesn't matter. The podman details don't matter: this is in the software release and people can always email you}

We run open-weight models on a server with a Xeon Gold 6342 CPU, H100 GPUs, and 1TB RAM. We serve models locally with vLLM~\citep{woosuk:vllm}, with a 128K context length. We use the official FP8 checkpoint for Qwen~3.6. We use the model's recommended sampling parameters. Gemma~4:
temperature~1.0, top-$p$~0.95, top-$k$~64; Qwen~3.6: temperature~1.0,
top-$p$~0.95, top-$k$~20. We impose a 6K token reasoning budget on both models. For \glm, we use the Hugging Face Inference API with temperature~1.0, top-$p$~0.95, high-effort reasoning, and a maximum output length of 11{,}000 tokens per turn (covering both reasoning and visible output).
Finally, we run \gptFull with `medium' reasoning, and run \sonnetFull with reasoning enabled and a 10K token reasoning budget. Reproducing our full experiment (400 tasks per model) requires approximately 28 GPU-hours for \gemma and 19 GPU-hours for \qwen on H100s if run sequentially (request-level concurrency can reduce this substantially), and approximately \$100 for \gpt, \$150 for \sonnet, and \$25 for \glm at standard API rates as of July 2026.

\section{System Prompt}
\label{appendix:prompts}

What follows is the template for the agent's prompt. We substitute the template variables (italicized) with the problem text and the language configuration.

\begin{transcript}

\system  You are a competitive programming solver that can interact with a computer shell to solve programming tasks.

    You are operating in an environment where you will think and issue AT LEAST ONE command, the system executes the command(s) in a subshell, you see the result, then you think and issue your next command(s).
    
    **CRITICAL REQUIREMENTS:**
    \begin{itemize}[nosep]
    \item Your response SHOULD include reasoning text explaining what you're doing
    \item Your response MUST include AT LEAST ONE bash tool call. You can make MULTIPLE tool calls in a single response when the commands are independent (e.g., reading multiple files).
    \end{itemize}
    
    \noindent \# Task Instructions
    
    \noindent \#\# Environment

    \begin{itemize}[nosep]
    \item Your working directory contains a tests/ directory, with test cases (0-indexed)
    \item the input is tests/input\_N.txt and the expected output is tests/output\_N.txt
    \item The problem statement is provided below in this message.
    \item Write your solution to: \emph{FILENAME}
    \item Your solution will be compiled and run as: \emph{COMPILE\_AND\_RUN}
    \item A test runner is provided at ./test.sh. It runs the solution against every visible test, diffs the output, and prints PASS or FAIL with each test input/output filename.
    \item To test your program, run exactly:
        ./test.sh
    \item These are the only tests I care about. If ./test.sh exits with status 0, the solution passes the tests; empty diff output means success.

    \end{itemize}

    \noindent \#\# Submission
    
    When ./test.sh exits with status 0, immediately submit by running:
        echo COMPLETE\_TASK\_AND\_SUBMIT\_FINAL\_OUTPUT
\user  Solve the following competitive programming problem in \emph{PROGRAMMING\_LANGUAGE}.

    \noindent\# Programming Problem:

    \emph{PROBLEM}

\end{transcript}

\section{Do Agents Cheat on MiniLCB?}
\label{appendix:cheating}

A natural concern with visible tests is that agents may cheat by hard-coding expected outputs rather than solving the problem.  To check for this, we use an LLM judge to flag suspicious solutions from the problem and final code, and we manually inspect every flagged case.  Across all problems, languages, and models we find only two instances, both from \gpt on the same problem instance: the model derives a nearly-correct rule, hits one visible test that contradicts it, and adds an exact-match guard for that case.  Cheating is therefore negligible and does not affect our analysis.

\section{Proprietary Model Results}
\label{appendix:proprietary}

\Cref{tab:mixed_effects_proprietary} reports the mixed-effects model for the proprietary models. As with the open-weight models, every language incurs a significant output-token cost penalty relative to Python after controlling for problem difficulty. 

\begin{figure*}[t]
    \centering
    \begin{subfigure}[t]{0.47\textwidth}
        \centering
        \begin{tikzpicture}[
  transition node/.style={
    circle,
    minimum size=0.82cm,
    align=center,
    text=white,
    font=\tiny,
    inner sep=0pt,
    outer sep=2pt
  },
  terminal node/.style={
    rounded corners=2pt,
    minimum width=0.68cm,
    minimum height=0.52cm,
    align=center,
    text=white,
    font=\tiny,
    inner sep=1pt,
    outer sep=2pt,
    draw=slate edge
  },
  transition/.style={
    -{Stealth[length=1.9mm,width=1.4mm]},
    line cap=round,
    line join=round,
    shorten >=1pt,
    shorten <=1pt
  },
  edge count/.style={
    pos=0.78,
    fill=white,
    fill opacity=0.72,
    text opacity=1,
    inner sep=0.5pt,
    font=\tiny
  },
  edge count slate/.style={edge count, text=slate edge},
  edge count stuck/.style={edge count, text=stuck fill},
  edge count improved/.style={edge count, text=improved fill},
  edge count breakthrough/.style={edge count, text=breakthrough fill},
  edge count stay/.style={edge count, text=stay fill},
  node distance=1.05cm and 1.12cm
]
\definecolor{start fill}{RGB}{151,166,186}
\definecolor{slate edge}{RGB}{97,111,132}
\definecolor{improved fill}{RGB}{44,151,229}
\definecolor{stuck fill}{RGB}{255,156,11}
\definecolor{breakthrough fill}{RGB}{3,182,197}
\definecolor{stay fill}{RGB}{76,174,78}

\node[terminal node, fill=start fill] (start) {start};
\node[transition node, fill=stuck fill, right=0.88cm of start] (stuck) {stuck};
\node[transition node, fill=improved fill, above=0.78cm of stuck] (improved) {improved};
\node[transition node, fill=breakthrough fill, below=0.78cm of stuck] (breakthrough) {break\\[-2pt]through};
\node[transition node, fill=stay fill, right=2.16cm of breakthrough] (stay) {stay};
\node[terminal node, fill=start fill, right=3.04cm of stuck] (submit) {submit};

\path[transition, slate edge]
  (start) edge[bend left=10] node[edge count slate, above, pos=0.5] {1} (improved)
  (start) edge node[edge count slate, above, pos=0.5] {2} (stuck)
  (start) edge[bend right=8] node[edge count slate, below, pos=0.5] {90} (breakthrough);

\path[transition, stuck fill]
  (stuck) edge[bend left=9] node[edge count stuck, left, pos=0.5] {1} (improved)
  (stuck) edge[bend right=10] node[edge count stuck, left, pos=0.5] {2} (breakthrough);

\path[transition, improved fill]
  (improved) edge[bend left=9] node[edge count improved, right, pos=0.5] {1} (stuck)
  (improved) edge[bend left=30] node[edge count improved, right] {1} (breakthrough);

\path[transition, breakthrough fill]
  (breakthrough) edge node[edge count breakthrough, above, pos=0.85] {1} (stay)
  (breakthrough) edge[bend left=5] node[edge count breakthrough, above] {92} (submit);

\path[transition, stay fill]
  (stay) edge[bend right=4] node[edge count stay, below, pos=0.6] {1} (submit);
\end{tikzpicture}
        \caption{\gpt Python}
    \end{subfigure}\hfill
    \begin{subfigure}[t]{0.47\textwidth}
        \centering
        \begin{tikzpicture}[
  transition node/.style={
    circle,
    minimum size=0.82cm,
    align=center,
    text=white,
    font=\tiny,
    inner sep=0pt,
    outer sep=2pt
  },
  terminal node/.style={
    rounded corners=2pt,
    minimum width=0.68cm,
    minimum height=0.52cm,
    align=center,
    text=white,
    font=\tiny,
    inner sep=1pt,
    outer sep=2pt,
    draw=slate edge
  },
  transition/.style={
    -{Stealth[length=1.9mm,width=1.4mm]},
    line cap=round,
    line join=round,
    shorten >=1pt,
    shorten <=1pt
  },
  edge count/.style={
    pos=0.78,
    fill=white,
    fill opacity=0.72,
    text opacity=1,
    inner sep=0.5pt,
    font=\tiny
  },
  edge count slate/.style={edge count, text=slate edge},
  edge count stuck/.style={edge count, text=stuck fill},
  edge count improved/.style={edge count, text=improved fill},
  edge count breakthrough/.style={edge count, text=breakthrough fill},
  edge count stay/.style={edge count, text=stay fill},
  node distance=1.05cm and 1.12cm
]
\definecolor{start fill}{RGB}{151,166,186}
\definecolor{slate edge}{RGB}{97,111,132}
\definecolor{improved fill}{RGB}{44,151,229}
\definecolor{stuck fill}{RGB}{255,156,11}
\definecolor{breakthrough fill}{RGB}{3,182,197}
\definecolor{stay fill}{RGB}{76,174,78}
\definecolor{limit fill}{RGB}{151,166,186}

\node[terminal node, fill=start fill] (start) {start};
\node[transition node, fill=stuck fill, right=0.88cm of start] (stuck) {stuck};
\node[transition node, fill=improved fill, above=0.78cm of stuck] (improved) {improved};
\node[transition node, fill=breakthrough fill, below=0.78cm of stuck] (breakthrough) {break\\[-2pt]through};
\node[terminal node, fill=limit fill, right=2.16cm of improved] (limit) {limit};
\node[terminal node, fill=start fill, right=3.04cm of stuck] (submit) {submit};

\path[transition, slate edge]
  (start) edge[bend left=10] node[edge count slate, above, pos=0.5] {4} (improved)
  (start) edge node[edge count slate, above, pos=0.5] {27} (stuck)
  (start) edge[bend right=8] node[edge count slate, below, pos=0.5] {61} (breakthrough);

\path[transition, stuck fill]
  (stuck) edge[out=140,in=220,looseness=6] node[edge count stuck, left] {13} (stuck)
  (stuck) edge[bend left=9] node[edge count stuck, left, pos=0.5] {1} (improved)
  (stuck) edge[bend right=10] node[edge count stuck, left, pos=0.5] {25} (breakthrough)
  (stuck) edge[bend left=4] node[edge count stuck, above] {1} (submit);

\path[transition, improved fill]
  (improved) edge[bend left=30] node[edge count improved, right] {3} (breakthrough)
  (improved) edge[bend left=9] node[edge count improved, above] {1} (submit)
  (improved) edge[bend left=16] node[edge count improved, below] {1} (limit);

\path[transition, breakthrough fill]
  (breakthrough) edge[bend left=5] node[edge count breakthrough, above] {89} (submit);
  
\end{tikzpicture}
        \caption{\gpt OCaml}
    \end{subfigure}
    \caption{Solution-snapshot transition graphs for \gpt, using the same node and edge semantics as \cref{fig:transition_grid}.}
    \label{fig:gpt-transitions}
\end{figure*}
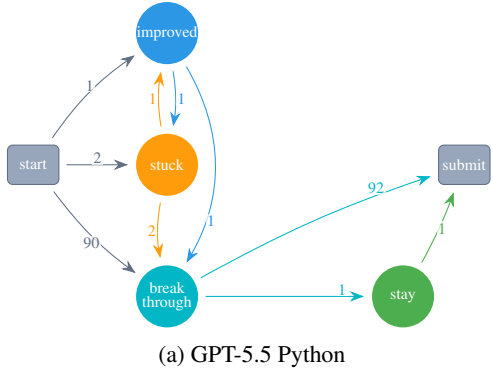
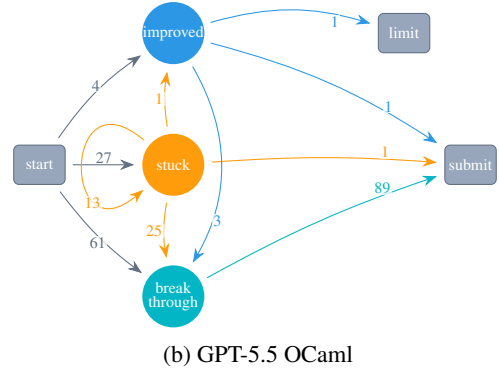

The higher token costs for non-Python languages are reflected in the trajectory patterns shown in \Cref{fig:gpt-transitions}, the \gpt counterpart to \Cref{fig:transition_grid} in the main text. 
On Python, \gpt almost always reaches a correct solution on the first write (\emph{start}$\to$\emph{breakthrough}: 90), with almost no \emph{stuck} activity.
On OCaml, it enters \emph{stuck} more often (\emph{start}$\to$\emph{stuck}: 27) and accumulates 13 stuck self-loops before recovering.

\Cref{tab:efficiency_proprietary,tab:gap-dist_proprietary} show that proprietary models follow a much more direct workflow than the open-weight models. Most trajectories consist of a single implementation phase followed by verification, indicating that the models typically write the solution once, immediately run \texttt{./test.sh}, and submit if all tests pass. Despite this streamlined execution, proprietary models still externalize substantial reasoning into code. For example, on \texttt{arc194\_a}, \gpt first writes a draft \texttt{solution.py} containing only commented-out derivations and a \texttt{pass}, then overwrites it with a complete implementation in the same turn. Similarly, \sonnet sometimes reasons in separate scratch files, solving one problem in a 700-line \texttt{think.py} before writing the final Rust solution. Proprietary models also frequently use Python as a reasoning language before translating to the target language. On \texttt{arc194\_e} in OCaml, \sonnet spends 23 turns exploring the problem in Python, writing a series of brute-force scripts to discover and verify the invariant before producing an OCaml solution that passes all 42 tests on the first write. Likewise, \gpt often embeds Python directly via \texttt{python3 -c} or heredocs to prototype algorithms or validate invariants, especially for OCaml (27\% of problems), Rust (19\%), and Java (18\%). These Python executions range from quick scratch calculations to full solution prototypes before translation. This behavior becomes more frequent on harder problems, suggesting that models adapt their workflow when additional algorithmic exploration is required.

\begin{table}[t]
\centering
\caption{Mixed-effects model results for \gpt and \sonnet: All three languages incur a significant output-token cost penalty relative to Python after controlling for problem difficulty.}
\label{tab:mixed_effects_proprietary}
\small
\setlength{\tabcolsep}{3pt}
\begin{tabular}{lrll}
\toprule
Contrast & \llap{Token-cost ratio} & 95\% CI & Significance \\
\midrule
\multicolumn{4}{l}{\textbf{\gpt} \textit{(ICC 0.94)}} \\
  Java vs.\ Python  & $1.30\times$ & $[1.18, 1.43]$ & $p < 0.001$ \\
  Rust vs.\ Python  & $1.36\times$ & $[1.24, 1.49]$ & $p < 0.001$ \\
  OCaml vs.\ Python & $1.30\times$ & $[1.19, 1.43]$ & $p < 0.001$ \\
\midrule
\multicolumn{4}{l}{\textbf{\sonnet} \textit{(ICC 0.97)}} \\
  Java vs.\ Python  & $1.18\times$ & $[1.08, 1.28]$ & $p < 0.001$ \\
  Rust vs.\ Python  & $1.16\times$ & $[1.07, 1.26]$ & $p < 0.001$ \\
  OCaml vs.\ Python & $1.28\times$ & $[1.18, 1.39]$ & $p < 0.001$ \\
\bottomrule
\end{tabular}
\end{table}

\begin{table}[t]
\centering
\caption{We run all intermediate solutions that the agent generates and report the number of tokens (turns) to the first solution that passes all tests. Proprietary models tend to submit their solution directly.}
\label{tab:efficiency_proprietary}
\small
\begin{tabular}{llrr}

\toprule
Model & Language &
 \makecell{Tokens (Turns) \\ to Solution} & \makecell{Tokens (Turns) \\ after Solution} \\
\midrule
\multirow{4}{*}{\gpt}
  & Python &   606  (2) &    42  (1) \\
  & Java   &   841  (2) &    42  (1) \\
  & Rust   &   782  (3) &    44  (1) \\
  & OCaml  &   910  (2) &    44  (1) \\
\midrule
\multirow{4}{*}{\sonnet}
  & Python &   878  (2) &   104  (1) \\
  & Java   &  1149  (1) &    96  (1) \\
  & Rust   &  1134  (1) &   136  (2) \\
  & OCaml  &  1484  (2) &   136  (1) \\
\bottomrule
\end{tabular}
\end{table}

\begin{table*}[t]
\centering
\footnotesize
\setlength{\tabcolsep}{1.8pt}
\renewcommand{\arraystretch}{1.12}
\newcommand{\gaphead}[1]{\tiny\bfseries\makecell[cc]{#1}}
\begin{tabular}{@{}>{\raggedright\arraybackslash}m{0.09\textwidth}>{\raggedright\arraybackslash}m{0.048\textwidth}*{12}{>{\centering\arraybackslash}m{0.057\textwidth}}@{}}
\toprule
\textbf{Model} & \textbf{Lang} &
\gaphead{Implement} &
\gaphead{Explore} &
\gaphead{Rewrite} &
\gaphead{Execute\\plan} &
\gaphead{Fix\\bug} &
\gaphead{Debug\\failure} &
\gaphead{Infra\\confusion} &
\gaphead{Revert} &
\gaphead{Verify} &
\gaphead{Edge\\case} &
\gaphead{Optimize\\perf.} &
\gaphead{Cosmetic\\refactor} \\
\midrule
  \textbf{\gpt} & java & 49\% & -- & -- & -- & 1\% & 1\% & -- & -- & 49\% & -- & -- & -- \\
   & ocaml & 42\% & 1\% & -- & <1\% & 6\% & <1\% & 10\% & -- & 40\% & -- & -- & <1\% \\
   & python & 48\% & -- & 1\% & 1\% & 1\% & -- & 1\% & -- & 48\% & -- & 1\% & -- \\
   & rust & 49\% & -- & 1\% & -- & 1\% & -- & -- & -- & 49\% & -- & -- & -- \\
\midrule
  \textbf{\sonnet} & java & 46\% & 1\% & 2\% & -- & 5\% & 1\% & -- & -- & 45\% & -- & -- & -- \\
   & ocaml & 45\% & 1\% & 2\% & -- & 7\% & <1\% & -- & -- & 43\% & -- & -- & <1\% \\
   & python & 46\% & 4\% & 1\% & 1\% & 2\% & 1\% & 1\% & -- & 42\% & -- & 4\% & -- \\
   & rust & 48\% & 1\% & 1\% & -- & 3\% & 1\% & -- & -- & 46\% & -- & -- & -- \\
\bottomrule
\end{tabular}
\caption{Percentage of \emph{spans} attributed to each label, for every (model,
  language) pair, showing how frequent each behavior is. \gpt and \sonnet are mostly \emph{implement}\,+\,\emph{verify}; debugging labels concentrate
  in OCaml; optimizations concentrate in Python.}
\label{tab:gap-dist_proprietary}
\end{table*}

\section{Agent-Observed vs.\ Retrospective Success}
\label{appendix:agent-observed}

Table~\ref{tab:efficiency} defines ``first solution'' by retrospectively replaying every intermediate snapshot against the full test suite: the agent may not have run this check itself. Our scaffold gives the agent an explicit mechanism to check its own solution (\texttt{test.sh}) and instructs it to run this check before submitting, so the efficient behavior, once a solution is correct, is to run the check and submit. Table~\ref{tab:efficiency-observed} recomputes Table~\ref{tab:efficiency} using ``first solution'' defined as the first snapshot at which the agent itself ran \texttt{test.sh} and observed \texttt{All tests passed.}, rather than a retrospective replay. Trajectories where the agent never itself observed a pass are excluded from this table, so $N$ is smaller than the number of solved trajectories.

\begin{table}[t]
\centering
\small
\caption{Table~\ref{tab:efficiency} recomputed with ``first solution'' defined as the first snapshot at which the agent itself ran \texttt{test.sh} and observed \texttt{All tests passed.} $N$ excludes trajectories where the agent never itself observed a pass.}
\label{tab:efficiency-observed}
\setlength{\tabcolsep}{3pt}
\begin{tabular}{@{}llrrr@{}}
\toprule
Model & Lang & $N$ & \makecell{Tokens (Turns) \\ to Solution} & \makecell{Tokens (Turns) \\ after Solution} \\
\midrule
\multirow{4}{*}{\gemma}
  & Python & 68 & 1512 (2) & 698 (1) \\
  & Java   & 81 & 2585 (2) & 3066 (1) \\
  & Rust   & 66 & 1736 (3) & 743 (1) \\
  & OCaml  & 57 & 2688 (4) & 743 (1) \\
\midrule
\multirow{4}{*}{\qwen}
  & Python & 82 & 1984 (4) & 86 (1) \\
  & Java   & 77 & 2390 (5) & 87 (1) \\
  & Rust   & 81 & 2930 (6) & 86 (1) \\
  & OCaml  & 71 & 3879 (10) & 82 (1) \\
\midrule
\multirow{4}{*}{\glm}
  & Python & 70 & 778 (2) & 79 (1) \\
  & Java   & 77 & 738 (2) & 74 (1) \\
  & Rust   & 75 & 1501 (3) & 66 (1) \\
  & OCaml  & 76 & 1448 (4) & 68 (1) \\
\bottomrule
\end{tabular}
\end{table}

The excluded trajectories are themselves informative. Table~\ref{tab:excluded-trajectories} reports, for every model and language, the number of such trajectories and the median tokens (turns) to the retrospectively passing snapshot and after it.

\begin{table}[t]
\centering
\small
\caption{Trajectories that are retrospectively solved but where the agent never itself observed a pass. $N$ is the number of such trajectories; medians are computed only over that $N$.}
\label{tab:excluded-trajectories}
\setlength{\tabcolsep}{3pt}
\begin{tabular}{@{}llrrr@{}}
\toprule
Model & Lang & $N$ & \makecell{Tokens (Turns) \\ to Snapshot} & \makecell{Tokens (Turns) \\ after Snapshot} \\
\midrule
\multirow{4}{*}{\gemma}
  & Python & 16 & 5797 (1)  & 45518 (10.5) \\
  & Java   & 3  & 8680 (1)  & 17970 (3) \\
  & Rust   & 20 & 6460 (1)  & 24768 (4.5) \\
  & OCaml  & 5  & 21627 (7) & 18909 (29) \\
\midrule
\multirow{4}{*}{\qwen}
  & Python & 0  & --- & --- \\
  & Java   & 1  & 28901 (9)  & 57042 (12) \\
  & Rust   & 0  & --- & --- \\
  & OCaml  & 3  & 46618 (36) & 924 (3) \\
\midrule
\multirow{4}{*}{\glm}
  & Python & 5  & 859 (1)  & 112 (1) \\
  & Java   & 0  & --- & --- \\
  & Rust   & 2  & 5588 (1) & 0 (0) \\
  & OCaml  & 0  & --- & --- \\
\bottomrule
\end{tabular}
\end{table}

For \gemma, of these trajectories, 43 of 44 (across all four languages) never ran \texttt{test.sh} at all---the agent is not checking repeatedly and failing, it is simply not checking. \gemma reaches a fully correct solution by a median of turn 1 in these cases, but then spends a median of 3--29 further turns on unrelated work. A simple stop-on-first-pass intervention would not by itself fix this: the agent is not verifying and stopping late, it is not verifying at all.

\section{Snapshot Span Label Distributions}

Tables~\ref{tab:stuck-breakthrough-labels} and~\ref{tab:stay-labels} give the full label breakdowns for \gemma\ discussed in \S\ref{sec:analysis:stuck} and \S\ref{sec:analysis:stay}. Table~\ref{tab:stuck-breakthrough-labels} shows \emph{fix bug} dominating OCaml's stuck$\to$breakthrough escapes, versus a spread across \emph{implement}, \emph{execute plan}, and \emph{optimize perf} on Python. Table~\ref{tab:stay-labels} shows \gemma\ spends turns looping on \textsc{stay} doing \emph{cosmetic refactor} and \emph{optimize perf}---unnecessary polishing under the benchmark's design.

\begin{table}[t]
\centering
\small
\begin{tabular}{@{}lrr@{}}
\toprule
\textbf{Label} & \textbf{Python} & \textbf{OCaml} \\
               & \small($n=12$)  & \small($n=53$) \\
\midrule
  \emph{implement}          & 17\% &  4\% \\
  \emph{rewrite}            &  --  &  4\% \\
  \emph{execute plan}      & 25\% &  6\% \\
  \emph{fix bug}           & 33\% & 60\% \\
  \emph{debug failure}     &  --  &  6\% \\
  \emph{infra confusion}   &  --  &  9\% \\
  \emph{optimize perf}     & 25\% &  --  \\
  \emph{cosmetic refactor} &  --  & 11\% \\
\bottomrule
\end{tabular}
\caption{Label distribution on edges from \emph{stuck} to \emph{breakthrough} for \gemma. OCaml breakthroughs are primarily due to \emph{fix bug}.}
\label{tab:stuck-breakthrough-labels}
\end{table}

\begin{table}[t]
  \centering
  \small
  \begin{tabular}{@{}lr@{}}
  \toprule
  \textbf{Label} & \textbf{\% of edges into \textsc{stay}} \\
  \midrule
  \textbf{\emph{cosmetic refactor}} & \textbf{34\%} \\
  \textbf{\emph{optimize perf}}     & \textbf{22\%} \\
  \emph{implement}                   & 18\% \\
  \emph{verify}                      &  8\% \\
    \emph{fix bug}                     &  5\% \\
  \emph{revert}                      &  4\% \\
  \emph{edge case}                   &  3\% \\
  \emph{explore}                     &  2\% \\
  \emph{debug failure}               &  2\% \\
  \emph{infra confusion}             &  1\% \\
  \emph{execute plan}                &  1\% \\
  \emph{rewrite}                     &  1\% \\
  \bottomrule
  \end{tabular}
  \caption{Label distribution across all edges entering \textsc{stay} for \gemma on Python ($n$=258). Over half are cosmetic refactoring or performance micro-optimization---changes that are unnecessary under the benchmark’s stated objective.}
  \label{tab:stay-labels}
  \end{table}

\begin{table*}[t]
\centering
\footnotesize
\setlength{\tabcolsep}{1.8pt}
\renewcommand{\arraystretch}{1.12}
\newcommand{\gaphead}[1]{\tiny\bfseries\makecell[cc]{#1}}
\begin{tabular}{@{}>{\raggedright\arraybackslash}m{0.09\textwidth}>{\raggedright\arraybackslash}m{0.048\textwidth}*{12}{>{\centering\arraybackslash}m{0.057\textwidth}}@{}}
\toprule
\textbf{Model} & \textbf{Lang} &
\gaphead{Implement} &
\gaphead{Explore} &
\gaphead{Rewrite} &
\gaphead{Execute\\plan} &
\gaphead{Fix\\bug} &
\gaphead{Debug\\failure} &
\gaphead{Infra\\confusion} &
\gaphead{Revert} &
\gaphead{Verify} &
\gaphead{Edge\\case} &
\gaphead{Optimize\\perf.} &
\gaphead{Cosmetic\\refactor} \\
\midrule
  \textbf{\gemma} & java & 31\% & 15\% & 8\% & <1\% & 8\% & 5\% & 2\% & 1\% & 15\% & 2\% & 5\% & 9\% \\
   & ocaml & 25\% & 16\% & 4\% & 2\% & 15\% & 10\% & 4\% & 12\% & 5\% & 2\% & 1\% & 4\% \\
   & python & 28\% & 13\% & 6\% & 2\% & 6\% & 2\% & <1\% & 5\% & 9\% & 2\% & 9\% & 19\% \\
   & rust & 27\% & 15\% & 5\% & 1\% & 6\% & 2\% & 1\% & 3\% & 17\% & 3\% & 5\% & 18\% \\
\midrule
  \textbf{\qwen} & java & 44\% & 21\% & 11\% & <1\% & 9\% & 12\% & <1\% & <1\% & 2\% & -- & <1\% & <1\% \\
   & ocaml & 36\% & 21\% & 7\% & <1\% & 19\% & 10\% & 4\% & -- & 2\% & -- & -- & 1\% \\
   & python & 49\% & 15\% & 9\% & <1\% & 12\% & 10\% & -- & -- & 2\% & -- & 2\% & 1\% \\
   & rust & 54\% & 18\% & 10\% & <1\% & 5\% & 7\% & <1\% & -- & 4\% & <1\% & <1\% & 1\% \\
\midrule
  \textbf{\glm} & java & 90\% & 5\% & -- & -- & -- & -- & -- & -- & 5\% & -- & -- & -- \\
   & ocaml & 77\% & -- & -- & -- & 16\% & <1\% & 2\% & -- & 5\% & -- & -- & <1\% \\
   & python & 92\% & -- & -- & -- & -- & -- & -- & -- & 8\% & -- & -- & -- \\
   & rust & 93\% & 1\% & -- & -- & 1\% & -- & -- & -- & 4\% & -- & -- & -- \\
\midrule
  \textbf{\gpt} & java & 97\% & -- & -- & -- & <1\% & <1\% & -- & -- & 2\% & -- & -- & -- \\
   & ocaml & 83\% & 3\% & -- & -- & 5\% & <1\% & 7\% & -- & 2\% & -- & -- & <1\% \\
   & python & 85\% & -- & 3\% & -- & <1\% & -- & 8\% & -- & 3\% & -- & 2\% & -- \\
   & rust & 95\% & -- & 3\% & -- & <1\% & -- & -- & -- & 2\% & -- & -- & -- \\
\midrule
  \textbf{\sonnet} & java & 69\% & 10\% & 5\% & -- & 8\% & 5\% & -- & -- & 3\% & -- & -- & -- \\
   & ocaml & 66\% & 15\% & 11\% & -- & 5\% & 1\% & -- & -- & 3\% & -- & -- & <1\% \\
   & python & 48\% & 32\% & 6\% & 1\% & 2\% & 6\% & 1\% & -- & 2\% & -- & 3\% & -- \\
   & rust & 65\% & 7\% & 8\% & -- & 4\% & 13\% & -- & -- & 3\% & -- & -- & -- \\
\bottomrule
\end{tabular}
\caption{Percentage of output tokens attributed to each label, for every (model, language) pair, showing how much each behavior accounts for the total token expenditure.}
\label{tab:gap-dist-tokens}
\end{table*}

\paragraph{Per-problem incidence and concentration.}
\label{appendix:per-problem}

In the trajectory diagrams, a pooled count sums over problems of very different lengths, so a handful of long trajectories can dominate it. For each behavior discussed in the main text, Tables~\ref{tab:per-problem-stuck}--\ref{tab:per-problem-poststuck} report, per model--language pair, an equal-weight rate (the average, across problems, of that problem's own local rate---edges or spans of the given type divided by that problem's total edges or spans), the percentage of problems in which the behavior occurs at least once, and the share of the pooled count contributed by only the five most-affected problems. Every behavior we highlight is broadly distributed across problems, not concentrated in a few long trajectories.

\emph{Stuck self-loops.} OCaml has both the highest equal-weight rate and problem incidence, and the lowest top-5 share, of the four languages, for both \gemma and \qwen (\Cref{tab:per-problem-stuck}).

\begin{table}[t]
\centering
\small
\caption{Stuck$\to$stuck self-loops: equal-weight rate, \% of problems affected, and top-5 problems' share of the pooled count. All values in \%.}
\label{tab:per-problem-stuck}
\begin{tabular}{@{}llrrr@{}}
\toprule
Model & Lang & Eq-wt rate & Problems & Top-5 share \\
\midrule
\multirow{4}{*}{\gemma}
  & python & 4.9  & 15 & 67 \\
  & java   & 6.3  & 19 & 55 \\
  & rust   & 5.7  & 15 & 69 \\
  & ocaml  & 46.6 & 71 & 16 \\
\midrule
\multirow{4}{*}{\qwen}
  & python & 2.8  & 8  & 81 \\
  & java   & 2.4  & 8  & 80 \\
  & rust   & 4.3  & 12 & 67 \\
  & ocaml  & 34.0 & 60 & 22 \\
\bottomrule
\end{tabular}
\end{table}

\emph{Stay self-loops.} Unlike stuck self-loops, this behavior is not OCaml-dominant: OCaml has the lowest equal-weight rate and problem incidence, and the highest top-5 share, of the four languages---this behavior is instead most prevalent on Python (\Cref{tab:per-problem-stay}), consistent with the case study in \S\ref{sec:analysis:stay}.

\begin{table}[t]
\centering
\small
\caption{Stay$\to$stay self-loops for \gemma: equal-weight rate, \% of problems affected, and top-5 problems' share of the pooled count. All values in \%.}
\label{tab:per-problem-stay}
\begin{tabular}{@{}llrrr@{}}
\toprule
Model & Lang & Eq-wt rate & Problems & Top-5 share \\
\midrule
\multirow{4}{*}{\gemma}
  & python & 18.1 & 34 & 33 \\
  & java   & 7.3  & 21 & 45 \\
  & rust   & 17.2 & 36 & 34 \\
  & ocaml  & 3.6  & 12 & 77 \\
\bottomrule
\end{tabular}
\end{table}

\emph{Error-recovery spans} (fix bug + debug failure + infra confusion + revert). OCaml has both the highest equal-weight rate and problem incidence, and the lowest top-5 share, of the four languages, for both \gemma and \qwen (\Cref{tab:per-problem-error-recovery}).

\begin{table}[t]
\centering
\small
\caption{Error-recovery spans (fix bug + debug failure + infra confusion + revert): equal-weight rate, \% of problems affected, and top-5 problems' share of the pooled count. All values in \%.}
\label{tab:per-problem-error-recovery}
\begin{tabular}{@{}llrrr@{}}
\toprule
Model & Lang & Eq-wt rate & Problems & Top-5 share \\
\midrule
\multirow{4}{*}{\gemma}
  & python & 7.2  & 31 & 40 \\
  & java   & 7.4  & 26 & 45 \\
  & rust   & 6.1  & 29 & 40 \\
  & ocaml  & 41.6 & 77 & 17 \\
\midrule
\multirow{4}{*}{\qwen}
  & python & 9.0  & 25 & 47 \\
  & java   & 12.6 & 29 & 41 \\
  & rust   & 11.9 & 32 & 42 \\
  & ocaml  & 38.3 & 71 & 21 \\
\bottomrule
\end{tabular}
\end{table}

\emph{Post-success spans} (optimize perf + cosmetic refactor). This behavior is Gemma-specific rather than OCaml-specific: \gemma's equal-weight rate and problem incidence exceed \qwen's in every one of the four languages, while \qwen shows almost none of this behavior anywhere (\Cref{tab:per-problem-poststuck}), consistent with \S\ref{sec:analysis:stay}.

\begin{table}[t]
\centering
\small
\caption{Post-success spans (optimize perf + cosmetic refactor): equal-weight rate, \% of problems affected, and top-5 problems' share of the pooled count. All values in \%.}
\label{tab:per-problem-poststuck}
\begin{tabular}{@{}llrrr@{}}
\toprule
Model & Lang & Eq-wt rate & Problems & Top-5 share \\
\midrule
\multirow{4}{*}{\gemma}
  & python & 16.6 & 45 & 31 \\
  & java   & 8.9  & 30 & 32 \\
  & rust   & 13.9 & 47 & 28 \\
  & ocaml  & 4.3  & 41 & 43 \\
\midrule
\multirow{4}{*}{\qwen}
  & python & 1.8  & 6  & 93 \\
  & java   & 0.3  & 2  & 100 \\
  & rust   & 2.0  & 7  & 75 \\
  & ocaml  & 1.8  & 12 & 54 \\
\bottomrule
\end{tabular}
\end{table}

\section{Measuring Agents' Token Usage}
\label{appendix:outputtoken}

For any nontrivial problem that takes more than one turn to solve, the agent's
conversation history is fed back into the model, turn after turn. So output
tokens become input tokens, and the total number of input tokens over several
turns can grow enormous. However, on a good inference platform, input processing
is much cheaper than generating outputs, especially when the platform supports
prefix caching~\citep{zheng2024sglang}. When an agent works autonomously to
solve a problem over several turns, prefix caching is very effective, and what
matters most is the time required to generate outputs. This is why, like many
others, we report the total number of output tokens generated over a single
conversation.

With contemporary reasoning models, the output tokens include both the reasoning
and normal output. In this paper, we work with both closed and open-weight
reasoning models. The closed models that we use, from OpenAI and Anthropic, do
not expose the reasoning output to users, but users are still charged for the
reasoning tokens produced. From the perspective of the user, output tokens cost
the same, whether or not the output is visible to the user.

Table~\ref{tab:cost-dispersion} reports median input tokens, reasoning tokens, and wall-clock time per trajectory. We do not report a monetary-cost breakdown, since dollar cost is a function of provider-specific pricing rather than an absolute, provider-agnostic quantity.

\section{Cost Dispersion Statistics}
\label{appendix:cost-dispersion}

Medians and the mixed-effects estimates describe a typical trajectory, but budgeting a workload depends on total cost, including the expensive tail and failed runs. \Cref{tab:cost-dispersion} reports, for every model and language, the ever-solved rate, mean and median output tokens, and tokens per success. The ever-solved rate is the share of trajectories that passed all tests at some snapshot, so it can exceed the final-solution accuracy in \Cref{fig:acccuray-vs-output-tokens}. Tokens per success is the total output tokens over all trajectories, failed ones included, divided by the number of ever-solved trajectories. For \gemma, OCaml is clearly the most expensive language to budget for: 45{,}390 tokens per success, 1.4--1.9$\times$ the other three languages. OCaml is also the most expensive for \qwen and \glm, but only narrowly, ahead of Java and Rust respectively. For \gpt and \sonnet, it is not the most expensive.

\begin{table*}[t]
\centering
\small
\caption{Cost dispersion by model and language. Mean, median, and tokens per success count output tokens. Input tokens, reasoning tokens, and wall-clock seconds are medians per trajectory.}
\label{tab:cost-dispersion}
\begin{tabular}{@{}llrrrrrrr@{}}
\toprule
Model & Lang & Ever solved & Mean tok. & Median tok. & Tok./success & Input tok. & Reasoning tok. & Wall-clock (s) \\
\midrule
\multirow{4}{*}{\gemma}
  & python & 84\% & 26{,}685 & 11{,}099 & 31{,}768 & 9{,}558  & 8{,}729  & 153 \\
  & java   & 82\% & 19{,}621 & 11{,}552 & 23{,}928 & 7{,}696  & 7{,}889  & 95 \\
  & rust   & 86\% & 25{,}103 & 15{,}014 & 29{,}190 & 13{,}378 & 10{,}429 & 162 \\
  & ocaml  & 62\% & 28{,}142 & 23{,}097 & 45{,}390 & 84{,}781 & 9{,}252  & 240 \\
\midrule
\multirow{4}{*}{\qwen}
  & python & 84\% & 18{,}786 & 2{,}830 & 22{,}451 & 12{,}092 & 1{,}920 & 25 \\
  & java   & 79\% & 21{,}300 & 3{,}357 & 26{,}832 & 14{,}034 & 2{,}329 & 33 \\
  & rust   & 84\% & 18{,}426 & 4{,}045 & 22{,}065 & 16{,}607 & 2{,}751 & 31 \\
  & ocaml  & 77\% & 21{,}002 & 8{,}546 & 27{,}331 & 43{,}058 & 4{,}407 & 63 \\
\midrule
\multirow{4}{*}{\glm}
  & python & 100\% & 1{,}979 & 861   & 1{,}979 & 5{,}280  & 592   & 13 \\
  & java   & 97\%  & 2{,}447 & 902   & 2{,}511 & 4{,}770  & 762   & 22 \\
  & rust   & 97\%  & 3{,}134 & 1{,}679 & 3{,}217 & 10{,}116 & 1{,}339 & 29 \\
  & ocaml  & 99\%  & 3{,}209 & 1{,}587 & 3{,}251 & 13{,}429 & 1{,}218 & 34 \\
\midrule
\multirow{4}{*}{\gpt}
  & python & 100\% & 1{,}998 & 643   & 1{,}998 & 4{,}770  & 244   & 18 \\
  & java   & 99\%  & 2{,}127 & 890   & 2{,}150 & 5{,}352  & 198   & 22 \\
  & rust   & 98\%  & 2{,}574 & 834   & 2{,}630 & 5{,}927  & 272   & 19 \\
  & ocaml  & 97\%  & 2{,}110 & 965   & 2{,}181 & 5{,}633  & 340   & 27 \\
\midrule
\multirow{4}{*}{\sonnet}
  & python & 92\% & 5{,}630 & 1{,}348 & 6{,}099 & 9{,}707  & 232   & 29 \\
  & java   & 96\% & 4{,}582 & 1{,}407 & 4{,}795 & 8{,}246  & 246   & 34 \\
  & rust   & 96\% & 4{,}333 & 1{,}324 & 4{,}534 & 11{,}214 & 224   & 30 \\
  & ocaml  & 95\% & 5{,}338 & 1{,}865 & 5{,}648 & 11{,}663 & 234   & 38 \\
\bottomrule
\end{tabular}
\end{table*}

\section{Trajectory Termination}
\label{appendix:termination}

\Cref{tab:termination} reports how each trajectory ends. Only \gemma and \qwen on OCaml exhaust the 40-turn budget in more than one trajectory, consistent with the stuck loops in \S\ref{sec:analysis:stuck}. Timeouts vary little across languages for every model. Memory is not a factor: across all 2{,}000 trajectories, only 6 contain a solution killed by the 256MB limit, one of them in OCaml, and none of these solutions would pass with more memory.

\begin{table}[t]
\centering
\small
\caption{How trajectories end, out of 100 per model and language: the agent submits, the trajectory times out, or the agent reaches the 40-turn limit.}
\label{tab:termination}
\setlength{\tabcolsep}{4pt}
\begin{tabular}{@{}llrrr@{}}
\toprule
Model & Lang & Submitted & Timeout & Turn limit \\
\midrule
\multirow{4}{*}{\gemma}
  & Python & 82 & 18 & 0 \\
  & Java   & 82 & 18 & 0 \\
  & Rust   & 78 & 22 & 0 \\
  & OCaml  & 49 & 29 & 22 \\
\midrule
\multirow{4}{*}{\qwen}
  & Python & 82 & 18 & 0 \\
  & Java   & 77 & 23 & 0 \\
  & Rust   & 81 & 19 & 0 \\
  & OCaml  & 71 & 18 & 11 \\
\midrule
\multirow{4}{*}{\glm}
  & Python & 74 & 26 & 0 \\
  & Java   & 77 & 23 & 0 \\
  & Rust   & 77 & 23 & 0 \\
  & OCaml  & 76 & 24 & 0 \\
\midrule
\multirow{4}{*}{\gpt}
  & Python & 93 & 7  & 0 \\
  & Java   & 94 & 6  & 0 \\
  & Rust   & 93 & 7  & 0 \\
  & OCaml  & 91 & 9  & 0 \\
\midrule
\multirow{4}{*}{\sonnet}
  & Python & 86 & 13 & 1 \\
  & Java   & 87 & 12 & 1 \\
  & Rust   & 87 & 13 & 0 \\
  & OCaml  & 87 & 12 & 1 \\
\bottomrule
\end{tabular}
\end{table}

\section{Representation vs.\ Development Cost}
\label{appendix:representation-cost}

To check whether the language penalty in \Cref{tab:mixed_effects} reflects verbose source code rather than agent behavior, we measured, for \gemma, \qwen, and \glm, the token cost of the final solution file in isolation, using a generic tokenizer (tiktoken \texttt{cl100k\_base}) as a cross-model proxy, since no single tokenizer spans all five models we evaluate. For each trajectory we compute a representation share, the final file's token count divided by the trajectory's total output tokens, and report the median of this per-trajectory ratio across problems---not the ratio of two separately-aggregated medians, which would conflate the two quantities.

\begin{table}[t]
\centering
\small
\caption{Median representation tokens (final solution file only), median total output tokens, and the median per-trajectory representation share, across \gemma, \qwen, and \glm trajectories.}
\label{tab:representation-cost}
\begin{tabular}{@{}lrrr@{}}
\toprule
Lang & Rep.\ tok. & Total tok. & Rep.\ share \\
\midrule
python & 190 & 2{,}548 & 5.8\% \\
java   & 276 & 2{,}972 & 8.4\% \\
rust   & 303 & 4{,}300 & 7.3\% \\
ocaml  & 272 & 6{,}706 & 4.5\% \\
\bottomrule
\end{tabular}
\end{table}

Across models, OCaml's median representation cost (272) is in line with the other compiled languages (Java 276, Rust 303), not with Python's lower 190---yet its median total is the largest of the four (6{,}706, roughly 2.6$\times$ Python's 2{,}548) and its representation share is the lowest. The extra cost is not representation. \Cref{tab:gap-dist-tokens} shows where it goes instead: models spend far more budget on error recovery in OCaml.

\section{Mixed-Effects Model Specification}
\label{appendix:mixed-effects-spec}

The mixed-effects model in \S\ref{sec:eval} is a linear mixed model fit to the natural log of output tokens:
\[
\log(\text{output\_tokens}) \sim \text{language} + (1 \mid \text{problem\_id})
\]
fit separately per model by restricted maximum likelihood (REML). Language is a fixed effect with Python as the reference level; problem identity is a random intercept, so the language coefficients reflect cost differences on the same problem. The response is log-transformed and the link function is the identity---this is a linear mixed model on the transformed response, not a generalized linear mixed model with a nonlinear link---so $\exp(\text{coef})$ recovers the multiplicative token-cost factor relative to Python. Budget-exhausted trajectories, which end in \texttt{timeout} or \texttt{turn\_limit}, enter the model with the output tokens recorded before termination; trajectories with zero recorded output tokens before the model times out are dropped.

Tables~\ref{tab:mixed_effects} and~\ref{tab:mixed_effects_proprietary} report 95\% confidence intervals for every coefficient alongside the point estimate.

\section{Use of AI Assistants}
Some code for this paper was written with AI assistants enabled.

\end{document}